\documentclass[preprint,useAMS,usenatbib]{aastex}

\usepackage{graphicx}
\usepackage{times}
\usepackage{amsmath}
\usepackage{epsf}
\usepackage{epstopdf}
\usepackage{hyperref}

\begin{document}


\title{A fast and portable Re-Implementation of Piskunov and Valenti's Optimal-Extraction 
Algorithm with improved Cosmic-Ray Removal and Optimal Sky Subtraction}
\author{A. Ritter}
\affil{National Central University}
\affil{300 Jhongda Rd, Jhongli City, Taoyuan County, Taiwan (R.O.C)}
\email{azuri.ritter@gmail.com}
\author{E. A. Hyde}
\affil{Department of Physics \& Astronomy, Macquarie University, Sydney, NSW 2109 Australia}
\and
\author{Q. A. Parker}
\affil{Department of Physics \& Astronomy, Macquarie University, Sydney, NSW 2109 Australia}
\affil{Research Centre for Astronomy, Astrophysics and Astrophotonics, Macquarie University, Sydney, NSW 2109 Australia}
\affil{Australian Astronomical Observatory, PO Box 296, Epping, NSW 1710, Australia}

\date{Accepted Nov. 26 2013}

\maketitle

\label{firstpage}

  We present a fast and portable re-implementation of Piskunov and Valenti's optimal-extraction algorithm 
  \citep{Piskunov2002} in \texttt{C/C++} together with full uncertainty propagation, improved cosmic-ray removal, 
  and an optimal background-subtraction algorithm. 
  This re-implementation can be used with IRAF and most existing data-reduction packages and leads to signal-to-noise 
  ratios close to the Poisson limit. The algorithm is very stable, operates on spectra from a wide range of instruments 
  (slit spectra and fibre feeds), and has been extensively tested for VLT/UVES, ESO/CES, ESO/FEROS, NTT/EMMI, NOT/ALFOSC, 
  STELLA/SES, SSO/WiFeS, and finally, P60/SEDM-IFU data.


\section{Introduction}
The concept of a variance-weighted (optimal) extraction for spectroscopic data has been
developed over the last two and a half decades in a series of papers
\citep{Robertson1986,Horne1986,Marsh1989, Mukai1990,Vershueren1990,Kinney1991,Valdes1992,Hall1994,
Piskunov2002,Bolton2010,Sharp2010}.
The most critical step in this procedure is the construction of an accurate profile model for
the slit function (spatial profile perpendicular to the dispersion direction\footnote{Note 
that in this paper we assume that the dispersion direction is only slightly tilted/curved 
with respect to the columns of the CCD, and the spatial profile is the transcept across a 
spectrum perpendicular to the dispersion direction (one resolution element
affects only one CCD row).}) for which several 
methods have been devised. While \citet{Horne1986} and
\citet{Marsh1989} describe methods of fitting low-order polynomials to straight and curved
orders, the spectrum-extraction algorithms by \citet{Valdes1992} use averaging of nearby
profiles. \citet{Kinney1991} fit binned, normalised profiles with high order polynomials to
generate profile models. Pis\-kunov and Valenti (\citealp{Piskunov2002}, hereafter P\&V) again
use an iteration algorithm to find the most probable two-dimensional spatial profile. Bolton \& Schlegel 
(\citealp{Bolton2010}) introduced a 2D Point-Spread Function (PSF) deconvolution model, but
state themselves that no computer in the near future will be able to solve the equations 
using a brute-force approach. 
Sharp \& Birchall (\citealp{Sharp2010}) again present an optimal-extraction algorithm for 
multi-object fibre spectroscopy assuming Gaussian spatial profiles based on the 2dF+AAOmega
multi-object fibre spectroscopy system \citep{Smith2004} on the Anglo-Australian Telescope (AAT).\\
Traditional sky-subtraction algorithms for slit spectra only look at the ends of the slit to
determine the underlying sky. For short slits, extended objects, or objects positioned at one
end of the slit, this is obviously problematic as the sky background might only be measurable 
in very few pixels or only one wing of the spatial profile. More sophisticated methods (e.g.\
\citealp{Sembach1996}, \citealp{Glazebrook2001}) have been developed to handle these problems,
but they come with huge costs in terms of observing time. Here we present a new optimal 
background-subtraction algorithm that avoids all these problems and delivers results close to the 
Poisson limit.\\
If the spatial profile or 2D PSF is not known a-priori, variance-weighted extraction only 
makes sense if the profile goes down to zero at the borders. The reason for this is the 
integral-normalisation of the (spatial) profile for each CCD row, hence a proper 
background subtraction is vital for the success of the determination
of the spatial profile. 
%
%
%
If the scattered light has not been subtracted
accurately or if the profile width is not set wide enough, the resulting profile image will
show jumps where the aperture borders cross a new column.\\
P\&V published their new and extremely powerful algorithms for reducing cross-dispersed echelle
spectra in 2002. Their Data-Reduction Pipeline (DRP) REDUCE was written in \texttt{IDL} and its only
purpose was to demonstrate the power of their new algorithms. They never intented to present a
stable pipeline which could easily be adopted to other instruments.
Unfortunately the \texttt{IDL}-based REDUCE package is slow and allows only experienced 
programmers to make use of it and adopt it to different instruments.\\
Here, we are presenting a re-implementation of the optimal-extraction algorithm from the REDUCE package
(status 10/09/2012) in \texttt{C/C++}. The original intention was to use P\&V's algorithm within
the Image Reduction and Analysis Facility (IRAF, \citealp{Tody1986}) as part of the automatic DRP \citep{Ritter2004} for the 
fibre-fed STELLA Echelle Spectrograph (SES, \citealp{Strassmeier2001}). As additional
features, we have introduced improved cosmic-ray and CCD-defect removal and new optimal background-subtraction algorithms. 
In the case of the background being nearly constant for each dispersion element, these new algorithms can determine 
the spatial profile AND the background (scattered light in case of fibre feeds, or 
scattered light + sky in case of slit spectra) in the same reduction step.
If the aperture 
definitions are provided in the form of an IRAF database file, these new programs can be 
called from any data-reduction package or programming language that supports the execution of 
\texttt{C/C++} code or external binaries. When run from within the STELLA-pipeline framework, 
all parameters are stored in a single parameter file, allowing for an easy overview and editing.\\
Our programs are freely available\footnote{\href{http://www.sourceforge.net/projects/stelladrp}{http://www.sourceforge.net/projects/stelladrp}} 
(including all used libraries), fast, portable,
extremely stable, and have been extensively bug fixed and tested for a wide range of
spectrographs, including the European Southern Observatory (ESO) Very Large Telescope (VLT) 
Ultraviolet and Visual Echelle (slit) Spectrograph (UVES, \citealt{Dekker2000}), ESO/Coude (fibre-fed) 
Echelle Spectrometer (CES, \citealt{Enard1982}), ESO/Fiber-fed Extended Range Optical (Echelle) Spectrograph (FEROS, 
\citealt{Kaufer1999}), ESO New Tech\-no\-lo\-gy Telescope (NTT)/ESO Multi-Mode Instrument (EMMI, \citealt{Dekker1986}), 
Northern Optical Telescope (NOT)/Andalucia Faint Object (slit) Spectrograph and Camera (ALFOSC), 
STELLA/SES, Siding Spring Observatory (SSO) Australian National University (ANU) 
Wide Field Spectrograph (WiFeS, \citealt{Dopita2010}) slitlet-based Integral Field Unit (IFU), and finally, 
Palomar Observatory 60inch Telescope (P60)/Spectral Energy Distribution (SED)-Machine 
(\citealt{Ben-Ami2012}) lenslet-based IFU data.\\
This paper is organised as follows: In Chapter 2 we will introduce our new algorithms; Chapter 3 is 
dedicated to the test results, and the conclusions and outlook to future work are presented in 
Chapter 4.\\

\section{Method}
\label{sec:method}
\subsection{Review of the original algorithm by P\&V}
P\&V (please refere to the original paper for a full explanation of the algorithm) calculate 
the most probable spatial profile by solving the problem
\begin{equation}
 \mathit{F} = \sum_{x,\lambda}\biggl[S_\lambda\sum_j{W_{x\lambda}^j P_{x\lambda}^j - D_{x\lambda}}\biggr]^2 + \Lambda\sum_{j}{(P_{x\lambda}^{j+1} - P_{x\lambda}^j)^2} = \mathrm{minimum},
 \label{PandV}
\end{equation}
where $x$ is the CCD column number of one spectral aperture row, $\lambda$ is the wavelength (or CCD row), 
$S_\lambda$ is the object spectrum, $P_{x\lambda}^j$ is the oversampled spatial 
profile, and $D_{x\lambda}$ is the flat-fielded and bias- and scattered-light subtracted CCD
aperture. The last term smooths the spatial profile, restricting
point-to-point variations, even if the order geometry does not constrain every point in $P$.
Given an oversampling factor $O$, the subpixel weights $W_{x\lambda}^j$ at a given wavelength $\lambda$ (CCD row $y$) for 
the subpixel $j$ are:
\begin{equation}
 W_{x\lambda}^j =
  \begin{cases}
   0 & j < j_0 \\
   \mathbf{FRACT}(x_0(\lambda)O) & j = j_0\\
   1/O & j=j_0+1,\dotsc,j_0+O-1\\
   1-\mathbf{FRACT}(x_0(\lambda)O) & j=j_0+O\\
   0 & j > j_0+O\\
  \end{cases}
\end{equation}
where $j_0$ is the index of the first subpixel that (partially) falls in detector pixel $[x,\lambda]$, and 
$\mathbf{FRACT}$ is the fraction of that first subpixel that is contained in the detector pixel. This structure is 
exactly the same for each pixel $x$ in a given CCD row.\\
Minimization of Eq.~\ref{PandV} produces two systems of linear equations:
\begin{equation}
 \sum_j(\mathbf{A}_{jk} + \Lambda\cdot\mathbf{B}_{jk})P_{x\lambda}^j = \mathbf{R}_k
\end{equation}
\begin{equation}
 S_\lambda = \mathbf{C}_\lambda / \mathbf{E}_\lambda
 \label{eq:pandvspec}
\end{equation}
with matrices given by:
\begin{equation}
 \mathbf{A}_{jk} = \sum_{x,\lambda}S_\lambda^2 W_{x\lambda}^j W_{x\lambda}^k
\end{equation}
\begin{equation}
 \mathbf{C}_{\lambda} = \sum_{x}D_{x\lambda}\sum_{j}P_{x\lambda}^{j} W_{x\lambda}^j
\end{equation}
\begin{equation}
 \mathbf{E}_{\lambda} = \sum_{x}\biggl[\sum_{j}P_{x\lambda}^{j} W_{x\lambda}^j\biggr]^2
\end{equation}
\begin{equation}
 \mathbf{R}_{k} = \sum_{x,\lambda}D_{x\lambda}S_\lambda W_{x\lambda}^k
\end{equation}
$\mathbf{B}_{jk}$ is the tri-diagonal matrix with $-1$ on both subdiagonals and 2 on the main diagonal,
except for the upper left and the bottom right corners, which contain 1. In the matrix equations above, the 
generic nomenclature $N=N_x\cdot O$ is used.\\
Cosmic-ray hits and CCD defects are automatically detected and marked as such in a mask $M_{x\lambda}$:
\begin{equation}
 M_{x\lambda} =
  \begin{cases}
   0 & \mathrm{if} (D_{x\lambda} - P_{x\lambda}\times S_\lambda)^2 > \sigma_{\mathrm{clip}}^2 \\
   1 & \mathrm{otherwise}\\
  \end{cases} 
\end{equation}
where $\sigma_{\mathrm{clip}}$ is determined from the whole swath in the first iteration, and from the individual
row in successive iteration steps.\\
The original programs by P\&V assumed that the scattered light was properly removed and the 
trace functions of the spectra/apertures were precisely known. Sky subtraction was then done 
by subtracting the median value of the lowest pixel values of one aperture row before the 
start of the calculation of the aperture profile. This works quite well and can even take 
care of (residual) scattered light if the slit is long and the 
observed star is close to the center of the slit. However, this procedure can be based on very
few pixel values and fails to deliver
optimal results for short slits or if the star is close to a slit edge, or if the traced 
aperture position is off by even a small percentage of a pixel.
To avoid these problems alternative methods for the background subtraction and the 
final extraction were developed and are presented in the following sections.\\

\subsection{Introduction of our new algorithms}
\begin{figure}[t]
  \plotone{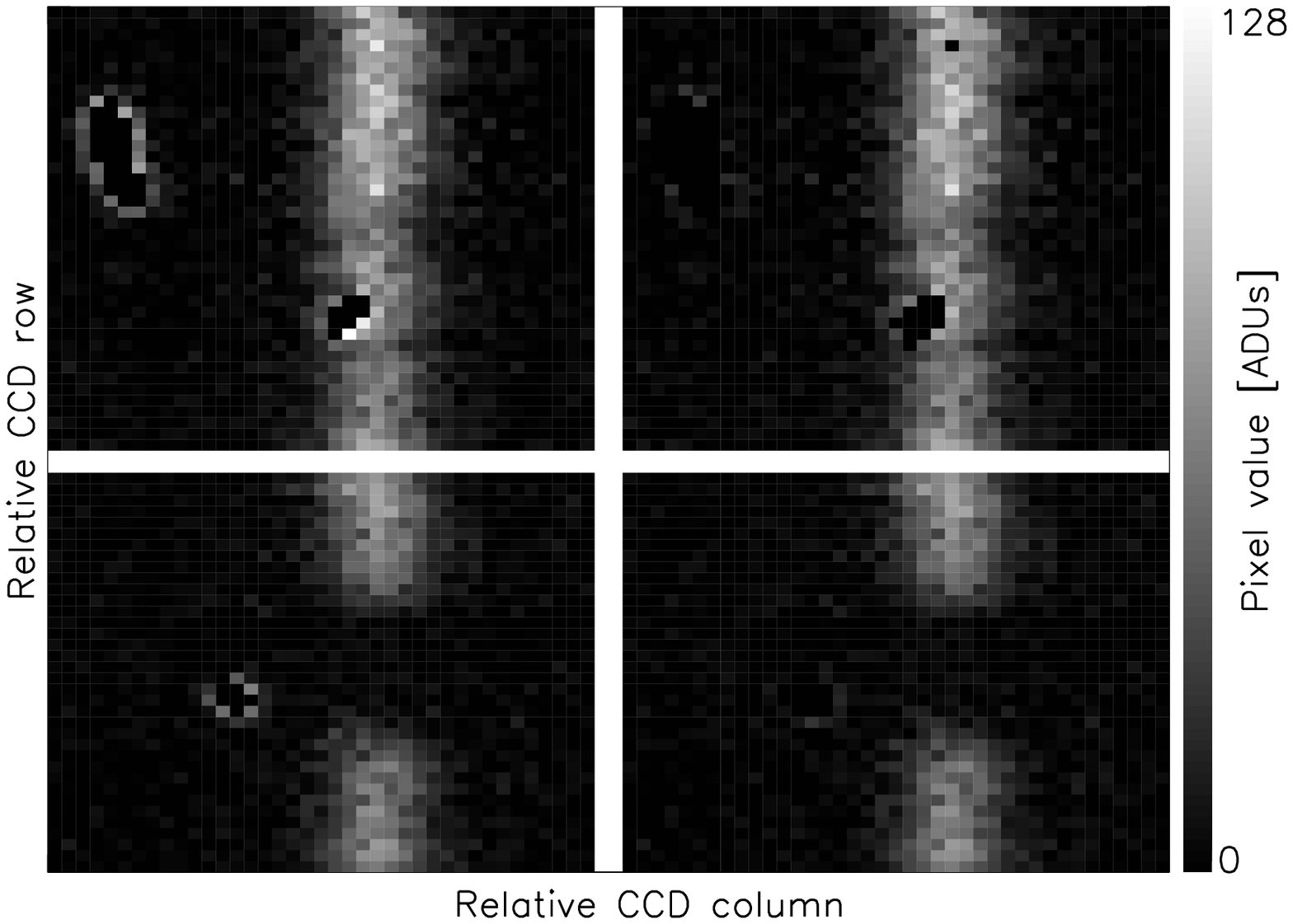}
  \caption{From left to right: Comparison of the original CCD image with cosmics-ray hits and sky line, the cosmic-ray removal of 
  LACosmics (still with sky line), of the REDUCE pipeline (identified pixels affected by cosmic rays set to zero and sky removed), 
  and our new algorithm implemented in the STELLA pipeline (same as for the REDUCE pipeline) for the two parts of the UVES spectrum 
  shown in Fig.~\ref{fig:ShadeSurf_CCD_profiles} which are affected by cosmic-ray hits.
  LACosmics does not completely remove the cosmic-ray hits and replaces the identified pixels with averages of the surrounding
  pixels. Both the REDUCE and the STELLA pipelines set the mask value of the identified pixels to zero and remove these pixels
  from the optimal extraction. While only the cores of the cosmic-ray hits are removed by the REDUCE pipeline, the STELLA pipeline 
  leaves no traces of them.}
\label{fig:cosmics}
\end{figure}

Optimal extraction is formally equivalent to scaling a known spatial profile across an object's 
spectrum to fit the sky-subtracted data. 
Here the latter approach is used to disentangle and extract the background (scattered 
light and/or sky) and the object spectrum in the same iterative procedure. An initial normalised 
spatial profile is determined first using only the rows with the highest maxima close to the 
peak of the integral-normalised spatial profile. 
Since adding a constant background to a profile results in lower maxima for the normalised
spatial profile, this procedure only takes into account rows that are not affected by sky
emission lines. After the initial determination of the spatial profile, the wings of the 
profile function are set to zero, removing background that affects all rows.
This is done by subtracting the minimum of the higher profile wing from the whole profile 
function, setting resulting negative values in the lower wing to zero, and subsequent 
re-normalisation of the whole profile. 
For each row, the most probable values for the background and the object spectrum are
then calculated by a weighted least-squares linear fit of the constant background and a scaled
spatial profile to the data (See \citealp{Bevington2003}):\\
\begin{equation}
 D_{x\lambda} = S_\lambda P_{x\lambda} + B_\lambda,
\end{equation}
where $B_\lambda$ is the (scattered light +) sky background.
Using the error estimates $\sigma_{x\lambda}$ from 
\begin{equation}
 \sigma_{x\lambda}^2 = \frac{D_{x\lambda}}{G} + \sigma_{RON}^2,
\end{equation}
where $D_{x\lambda}$ is the number of counts in pixel $[x,\lambda]$ after bias subtraction, $G$ is the detector 
gain, and $\sigma_{RON}$ is the read-out noise of the detector, the $\chi_\lambda^2$ to be minimised
for each resolution element $\lambda$ can be written as
\begin{equation}
 \chi_\lambda^2 = \sum_x{W_{x\lambda}[D_{x\lambda} - (S_\lambda P_{x\lambda} + B_\lambda)]^2}
\end{equation}
with the pixel weight
\begin{equation}
 W_{x\lambda} = \frac{1}{\sigma_{x\lambda}^2},
\end{equation}
where $\sigma_{x\lambda}$ is the uncertainty of pixel $x$ in a detector row ($y$ or $\lambda$) of the aperture.
When $\chi_\lambda^2$ is minimised, the estimated parameter values of the linear model can be computed
as
\begin{equation}
 S_\lambda = \frac{1}{\Delta_\lambda}\biggl(\sum_x{W_{x\lambda}}\sum_x{W_{x\lambda} P_{x\lambda} D_{x\lambda}} - \sum_x{W_{x\lambda} P_{x\lambda}}\sum_x{W_{x\lambda} D_{x\lambda}}\biggr),
\end{equation}
\begin{equation}
 B_\lambda = \frac{1}{\Delta_\lambda}\biggl(\sum_x{W_{x\lambda} P_{x\lambda}^2}\sum_x{W_{x\lambda} D_{x\lambda}} - \sum_x{W_{x\lambda} P_{x\lambda}}\sum_x{W_{x\lambda} P_{x\lambda} D_{x\lambda}}\biggr),
\end{equation}
where
\begin{equation}
 \Delta_\lambda = \sum_x{W_{x\lambda}}\sum_x{W_{x\lambda} P_{x\lambda}^2} - \biggl( \sum_x{W_{x\lambda} P_{x\lambda}} \biggr)^2
\end{equation}
The background is then subtracted 
from the input image and the spatial profile is re-calculated using all rows. This procedure
is repeated with the new profile until the spatial profiles $P_{x\lambda}$ converge. 
Convergence is assumed when
\begin{equation}
 \sum_{x,\lambda}\biggl(P_{x\lambda}^{old} - P_{x\lambda}^{new}\biggr)^2 < \frac{\overline{P_{x\lambda}^{new}}}{10^6},
\end{equation}
and normally reached 
within 10 iterations. Tests have shown that this is a good convergence criterion.
Negative values for the background are not allowed during the calculation
as they are not physical.\\
Another flaw in the original algorithm by P\&V was that problematic pixels were detected by
either looking at the whole swath or individual rows. The signal (and therefore the uncertainty per pixel) 
is a lot higher close to the center of the spatial profile $P_{x\lambda}$ compared to the wings of the profile. Therefore 
not all pixels affected by a cosmic-ray hit were identified, only the ones in the central area of the hit. In our new 
programs cosmic rays and bad pixels are identified and marked as such in a two step procedure. First we calculate the variance
per column instead of row. This ensures that each pixel is only compared to pixels with a similar signal strength and 
uncertainty. Setting the sigma-clipping factor high enough ($> 6$) makes sure that sky lines are not misidentified as problematic
pixels. Repeating this sigma-clipping procedure a number of times (which can be adjusted by the user) then also removes 
pixels which were affected by cosmic-ray hits but are not in the central region of the hit. The second step is done during 
the weighted linear fitting of the spatial profile to the CCD row containing the spectrum of the object and the background.
In this step pixels with 
\begin{equation}
 (D_{x\lambda} - P_{x\lambda}S_\lambda - B_\lambda)^2 > \sigma_{\mathrm{clip}}^2,
\end{equation}
are added to the problematic pixel mask by setting the mask value to zero.
Multiplying the CCD row 
and the spatial profile with the mask before the final fitting procedure then results in proper removal 
of the cosmic-ray hits/detector defects in the final extracted spectra (See Fig.~\ref{fig:cosmics}). In the case one prefers to 
reject cosmic rays before the optimal-extraction procedure with LACosmics in IRAF \citep{vanDokkum2001}, this is also supported 
by our STELLA pipeline. However, as shown in the same figure, at least the IRAF task 
\texttt{lacos\_spec}\footnote{\href{http://www.astro.yale.edu/dokkum/lacosmic/download.html}{http://www.astro.yale.edu/dokkum/lacosmic/download.html}}
can still leave traces of the cosmic-ray hit on the CCD.\\

The STELLA pipeline offers multiple possibilities for removing the scattered light. 
First of all, standard-IRAF's \texttt{apscatter} task from the \texttt{noao.twodspec.apextract} 
package is implemented in the STELLA DRP. Additionally, we re-implemented 
P\&V's scattered-light 
subtraction algorithm, as well as the Kriging algorithm (\citealt{b2}, \citealt{b3}). Kriging 
is very successfully used in Geostatistics for predicting sedimentary rock layers from a small number of
bore holes, but is very expensive in terms of computing time. However, clustering the scattered 
light leads to good results in an acceptable time frame. For the clustering two versions are implemented. 
The first version assigns the minimum of each rectangular area (cluster) on the CCD (size defined by the user) 
as scattered-light value to the center of the cluster. This version is very useful if the scattered light needs to 
be subtracted before the spectral apertures are identified. The second version only looks at areas on the CCD which 
are not part of a spectral aperture, and assigns the mean value of each cluster (again the sizes of clusters are defined
by the user) to the center of each cluster. The complete algorithm will be described in detail 
in a subsequent paper. Last but not least, setting the wings of the spatial profile (as described 
in the next paragraphs), can in principle also take care of scattered light, at least if the scattered
light is nearly constant underneath a given CCD row.\\
In addition to the (bug-fixed) optimal-extraction algorithm described by P\&V (with and without 
sky subtraction), the STELLA pipeline also offers two new algorithms for the sky subtraction and
the optimal extraction of the object spectrum in the same iterative procedure. In the first 
version, the initial spatial profile is calculated only from rows which are not affected by sky
lines. The second version also sets the wings of the spatial profile to zero, removing sky 
continuum and (residual) scattered light as well.\\
Our new algorithms do not involve any re-binning and therefore leave the resulting extracted spectra
uncorrelated. Note however that they are still limited to sky lines which are aligned with the
rows/columns of the CCD\@. For the final extraction step, the (oversampled) spatial profile can
be cross-correlated to the CCD row in order to find and eliminate offsets between the trace 
function and the real position of the object spectrum on the CCD\@. After the cross-correlation for 
each row, the offsets are fit with a low-order polynomial before being added to the trace function.\\
All libraries we used for the re-implementation of P\&V's optimal-extraction algorithm
are open source and freely available. For reading and writing \texttt{fits} files the
\texttt{cfitsio} library has been included. Mathematical procedures use the \texttt{BLITZ++}
library \citep{Veldhuizen1997}, which provides performance on par with \texttt{FORTRAN} 77/90.
The \texttt{IDL} code from P\&V was ported as closely as possible to \texttt{C++}. Bugs
like possible divisions by zero, which caused the program to crash, were identified and removed
in our re-implementation.\\

\subsection{Error estimation and propagation}
In the REDUCE pipeline, the actual uncertainties of the individual pixel values are not taken into account 
at all during the extraction or for the final estimation of the uncertainties. 
As an error estimate of the final object spectrum P\&V calculate
\begin{equation}
 \sigma_{\lambda} = \frac{\sum_x{(D_{x\lambda} - S_\lambda P_{x\lambda})^2}}{\sum_x P_{x\lambda}-\overline{P_{x\lambda}}},
 \label{eq:reduce_sigma}
\end{equation}
where $\overline{P_{x\lambda}}$ is the mean value of the spatial profile for the CCD row $\lambda$ ($y$). 
In our re-implementation of the original algorithm, we calculate the error estimate for the final extracted object 
spectrum from Eq.~\ref{eq:pandvspec} and the uncertainties for each pixel in the CCD row:
\begin{equation}
 \sigma_{\lambda}^2 = \frac{\sum_x{\sigma_{x\lambda}^2 F_{x\lambda}^2}}{(\sum_x{F_{x\lambda}^2)^2}}
 \label{eq:sigma_orig}
\end{equation}
where 
\begin{equation}
 F_{x\lambda} = \sum_{j}P_x^{j} W_{x\lambda}^j
\end{equation}

The new algorithms presented here use the estimated pixel errors,
which are given to the program in an additional error image, to calculate the most probable
spectrum and background, and propagate the errors through the entire extraction process. This 
procedure results in much more realistic error estimates for the final object spectra (See Sec.~\ref{sec:TestResults}).
Assuming that flat-fielding and scattered-light subtraction do not introduce additional errors,
the uncertainties of the final spectrum and constant background
can be calculated as
\begin{equation}
 \sigma_{S,\lambda}^2 = \frac{1}{\Delta_\lambda}\sum_x{W_{x\lambda}}  \mathrm{~and}
 \label{eq:sigma_s_new}
\end{equation}
\begin{equation}
 \sigma_{B,\lambda}^2 = \frac{1}{\Delta_\lambda}\sum_x{W_{x\lambda} P_{x\lambda}^2}
 \label{eq:sigma_b_new}
\end{equation}
where $\sigma_{S,\lambda}$ and $\sigma_{B,\lambda}$ are the uncertainties for the object 
spectrum and the background, respectively.\\

\subsection{Data flow chart}
\begin{figure}
 \plotone{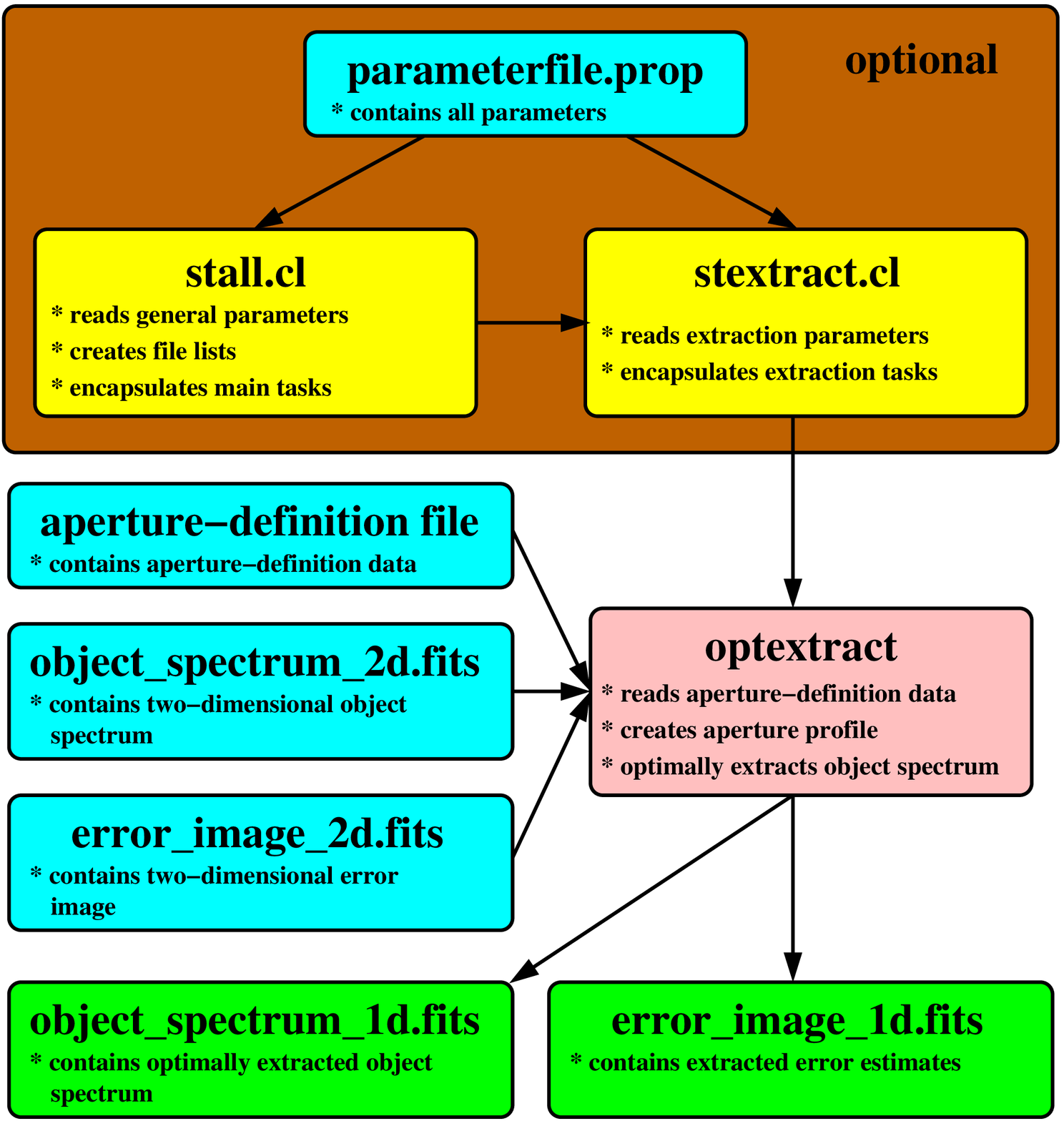}
  \caption{Flow chart of the parameters and data for our new \texttt{optextract} program,
  if invoked by the STELLA pipeline. The yellow boxes show STELLA tasks, the light blue boxes
  input files, and the green boxes the output files. Files within the brown box are optional
  for the execution of the task.}
\label{fig:FlowChart}
\end{figure}

If the new programs are invoked by the STELLA pipeline, all parameters for the optimal
extraction procedure are stored in one parameterfile. This provides for both an easy overview
of the parameters and their values, as well as easy editing. The flow chart for this case is
shown in Fig.~\ref{fig:FlowChart}.\\
Our new \texttt{C/C++} programs can be utilised to calculate the scattered light, the
normalised Flat, and the instrument profile (quantum efficiency and Blaze function) in the case that
a continuous light source over the whole spectral range is available (or the spectrum of the Flat 
field lamp is known), as well as to remove CCD defects and cosmic-ray hits,
and to optimally extract the target data and sky to one-dimensional spectra.
As an additional feature, they can create the profile image and the reconstructed
object and sky images in 2D, where always possible failures can be easily spotted by comparing
them to the original image.\\

\section{Test results}
\label{sec:TestResults}

\begin{figure}[t]
  \plotone{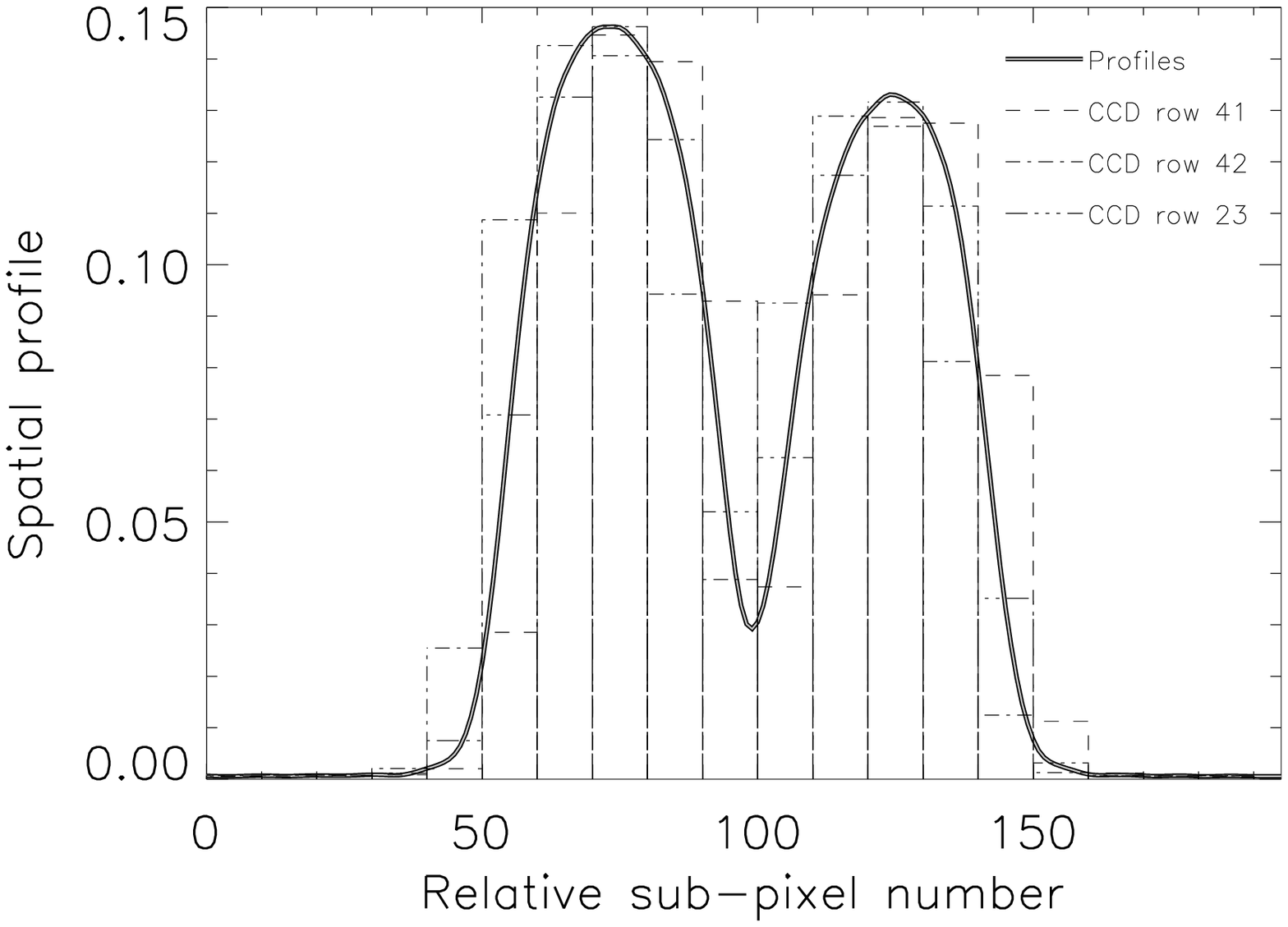}
  \caption{Comparison of the aperture profiles perpendicular to the dispersion axis for
  ESO/FEROS (with 2-slice image slicer). The thick black line shows the profile calculated with the \texttt{IDL}-REDUCE
  package from P\&V with 10 times oversampling. The over-plotted thin gray line shows the same
  profile calculated with the STELLA pipeline. Also shown are the histograms of 3 CCD rows in the same aperture: row 41 is a row just 
  before the aperture trace function crosses the border between two CCD columns, row 42 is the row just after that, and row 23 is a 
  row where the trace function goes through the center of the CCD column. All curves/histograms are normalised to an integral of
  one.}
  \label{fig:ProfileComparison}
\end{figure}

\begin{figure}[t]
  \plotone{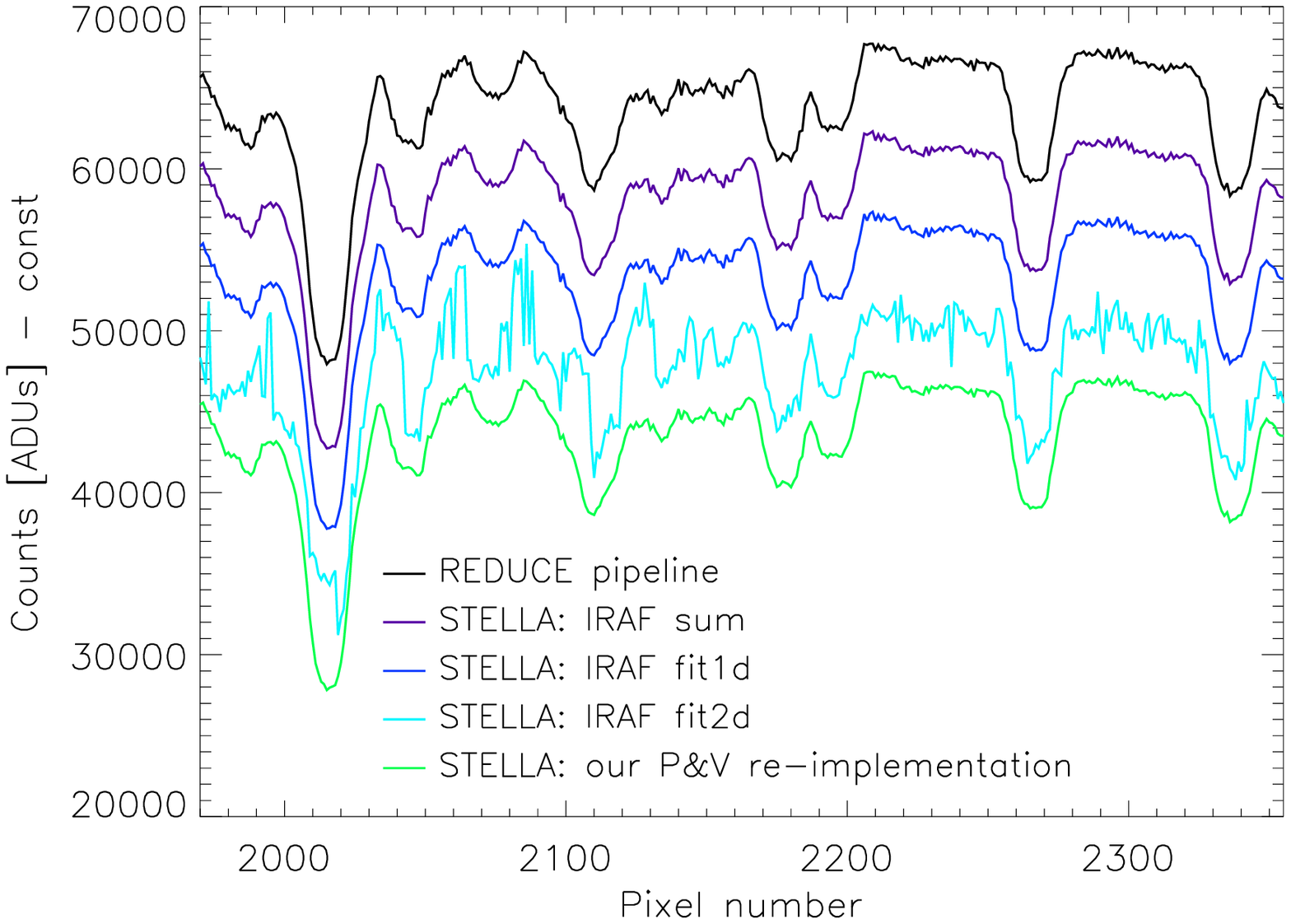}
  \caption{Comparison of the extraction algorithms from the REDUCE and STELLA pipelines for a part of one
  spectral order (5417--5622 \AA) of the variable K0V-star LQ Hya (HD 82558),
  observed with ESO/FEROS\@. The lower spectra were shifted downwards for better visibility.}
\label{fig:ComparisonREDUCE-STELLA}
\end{figure}

Our new \texttt{C/C++} modules/classes were tested using the `black box' testing strategy.
Testing modules were written to test the different constructors and procedures.\\
`Black box' means that the source code is not known. The tester only knows
the interface and checks the results of the procedures due to the possible entry points
and variable limits. The testing modules compare the results of the procedures to the
expected ones. If a single test goes wrong, the testing module returns immediately.\\
The comparison with the original IDL pipeline (REDUCE) by P\&V could not be done
completely, because REDUCE crashed. After fixing a few bugs, the optimal-extraction
algorithm was working properly for at least a few spectral orders of ESO/FEROS. As shown
in Fig.~\ref{fig:ProfileComparison}, the resulting profiles are indistinguishable.\\
The comparison of the original REDUCE pipeline without sky subtraction to the individual
extraction methods performed by the STELLA pipeline (IRAF sum, fit1d, fit2d, and our re-implementation of P\&V's algorithm) 
is shown in Fig.~\ref{fig:ComparisonREDUCE-STELLA}. While the IRAF fit2d algorithm completely fails, the other
extraction methods deliver comparable results. Note however that our re-implementation leads to the smoothest object spectrum.\\

\begin{figure*}
 \begin{center}
  \begin{minipage}[t]{0.999\textwidth}
    \hspace*{-1mm}
    a)\hspace*{-4mm}
    \begin{minipage}[t]{0.33\textwidth}
     \centerline{\includegraphics[width=\textwidth]{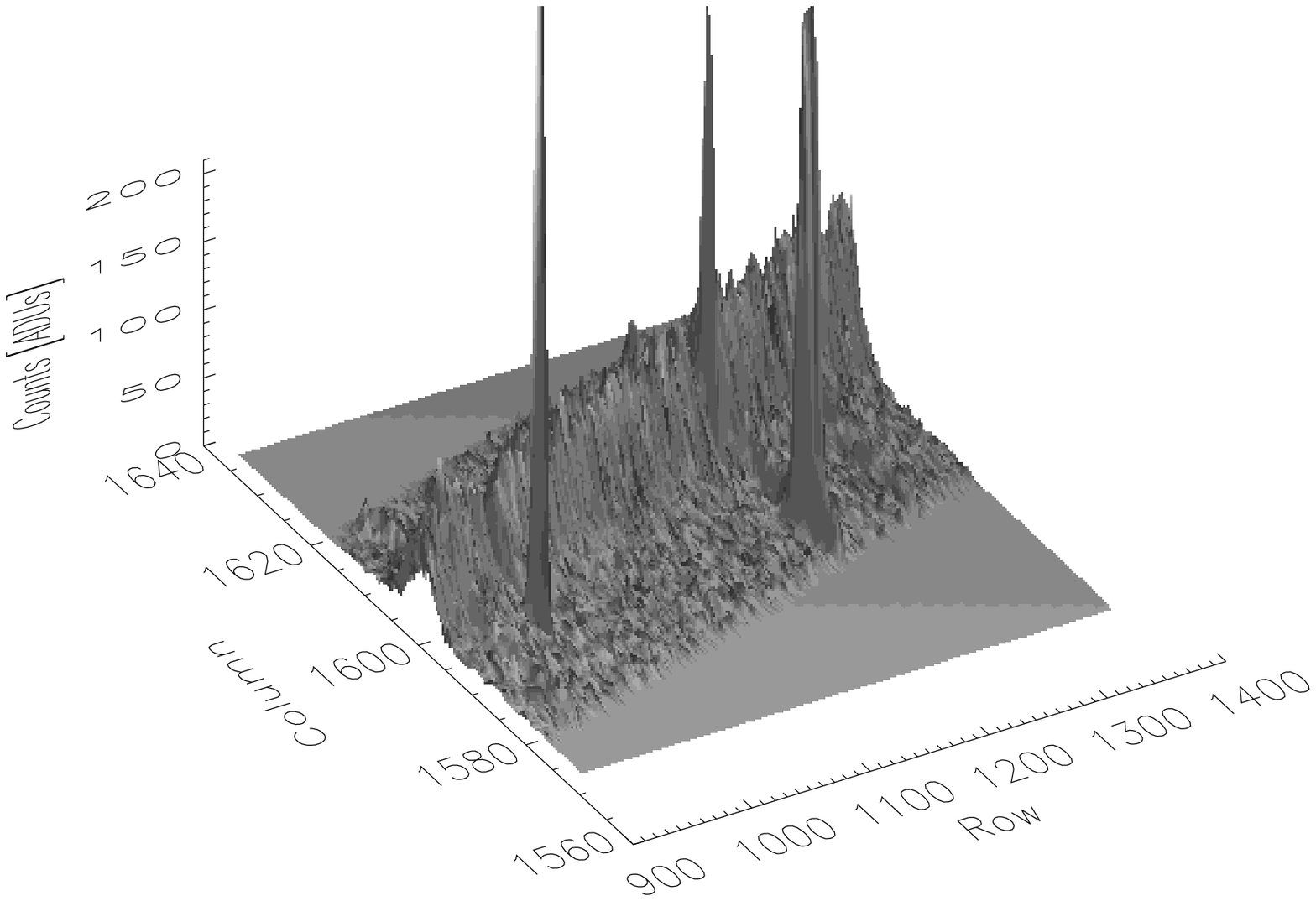}}
    \end{minipage}
    b)\hspace*{-4mm}
    \begin{minipage}[t]{0.33\textwidth}
     \centerline{\includegraphics[width=\textwidth]{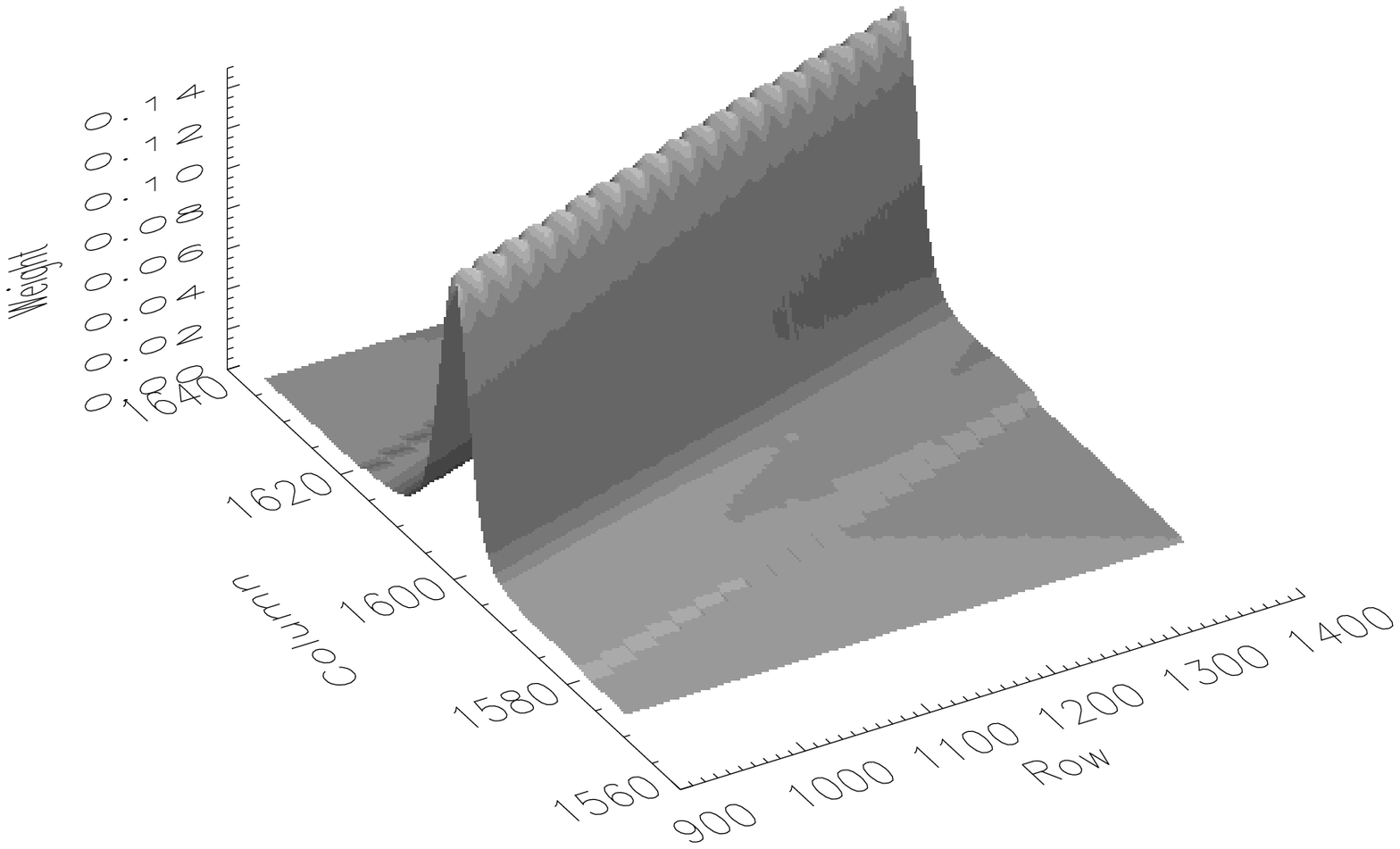}}
    \end{minipage}
    c)\hspace*{-4mm}
    \begin{minipage}[t]{0.33\textwidth}
     \centerline{\includegraphics[width=\textwidth]{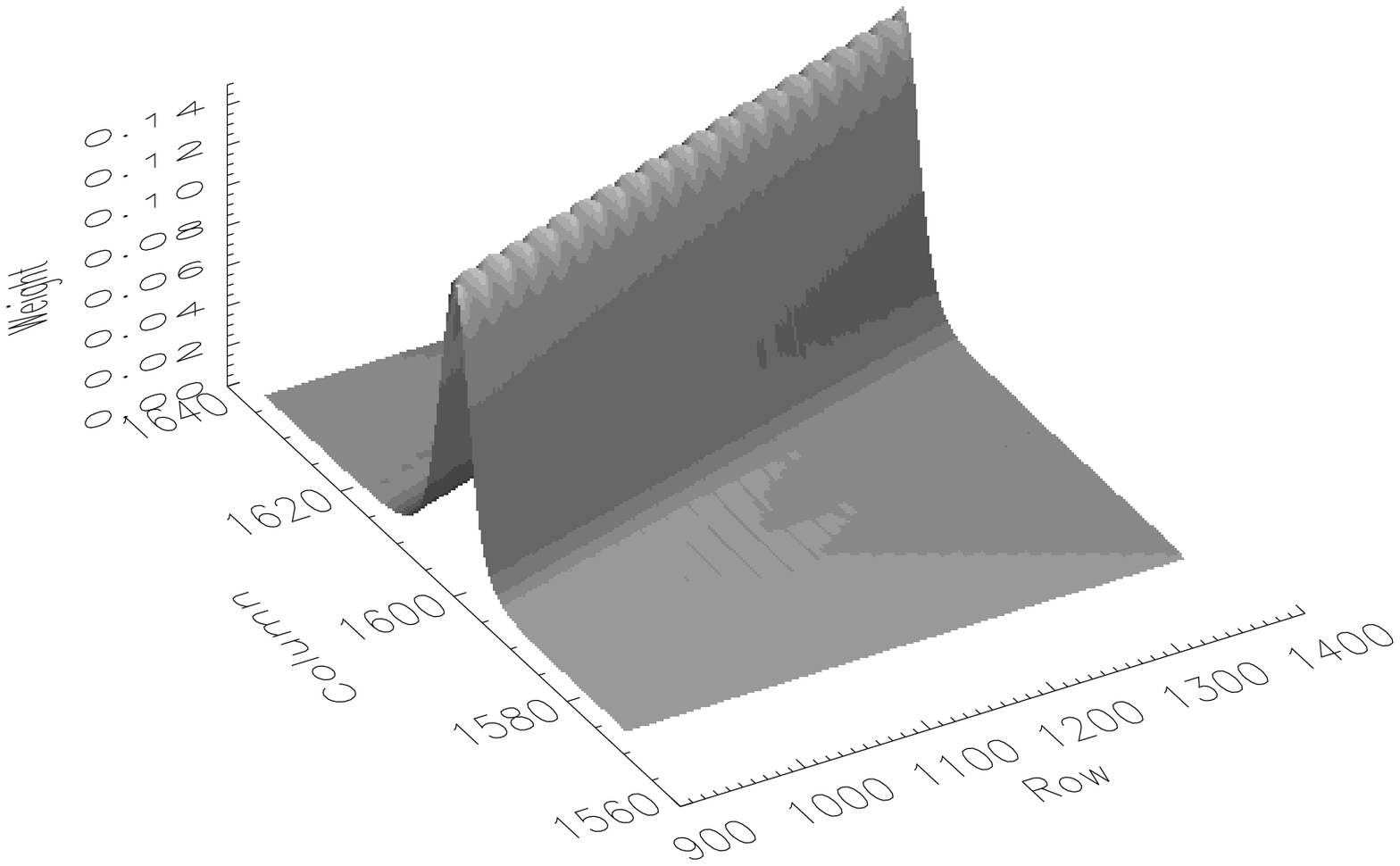}}
    \end{minipage}
  \end{minipage}
 \end{center}
 \caption{Part of a spectral order of LMC-X1 in the red, observed with UVES, showing strong sky lines as well as 
cosmic-ray hits. The spectral region chosen for this test example is $5886.8~\mathrm{\AA} < \lambda < 5896.5~\mathrm{\AA}$ 
(CCD rows 1363--914 respectively),
because it shows 3 sky-emission lines at 5888.192, 5889.959, and 5895.932 \AA~(CCD rows
1300, 1220, 940 respectively), 2 stellar absorption lines at $5890.30~\mathrm{\AA}$ and $5894.98~\mathrm{\AA}$~(CCD rows 
288 and 72), and 3 cosmic-ray hits,
one in the middle of the spatial profile, one in the wings of an absorption line,
and one in a sky line: a) Original
image after bias subtraction, flat-fielding, and scattered-light subtraction.\ b) Spatial profile calculated with our
re-implementation of the original algorithm by P\&V.\ c) Spatial profile calculated with our new algorithm. Note that 
the wings of the profile shown in b) show small steps at the limits of the extraction width, which disappeared in c). The 
ripples visible at the top of the profiles are caused by the trace function crossing columns of the CCD.}
\label{fig:ShadeSurf_CCD_profiles}
\end{figure*}

As a test spectrum for the comparison of our new algorithms to the REDUCE pipeline and the UVES pipeline 
\citep{Ballester2000} we chose a low-SNR observation of LMC-X1 (a stellar-mass black-hole candidate as
revealed by the properties of the surrounding material, \citealt{Wilms1998}) showing strong sky lines as 
well as cosmic-ray hits. The spectral region chosen for this test example is 
$5886.8~\mathrm{\AA} < \lambda < 5896.5~\mathrm{\AA}$ (CCD rows 1363--914 respectively),
because it shows 3 sky-emission lines at 5888.192, 5889.959, and 5895.932 \AA~(CCD rows
1300, 1220, and 940 respectively), 2 stellar absorption lines at $5890.30~\mathrm{\AA}$ and 
$5894.98~\mathrm{\AA}$~(CCD rows 288 and 72), and 3 cosmic-ray hits,
one in the middle of the spatial profile, one in the wings of an absorption line,
and one in a sky line. The mean value of the SNR in this region is approximately 20, 
making this a low-SNR spectrum where the optimal-extraction algorithm of the UVES pipeline should 
perform at its best. Fig.~\ref{fig:ShadeSurf_CCD_profiles} shows the original CCD section after bias 
subtraction, flat-fielding, and scattered-light subtraction as well as a comparison of the spatial profiles 
calculated with our new optimal sky- and object-extraction algorithms to the algorithm by P\&V. 

\begin{figure*}
 \begin{center}
  \begin{minipage}[t]{0.999\textwidth}
    \hspace*{-1mm}
    a)\hspace*{-4mm}
    \begin{minipage}[t]{0.33\textwidth}
      \centerline{\includegraphics[width=\textwidth]{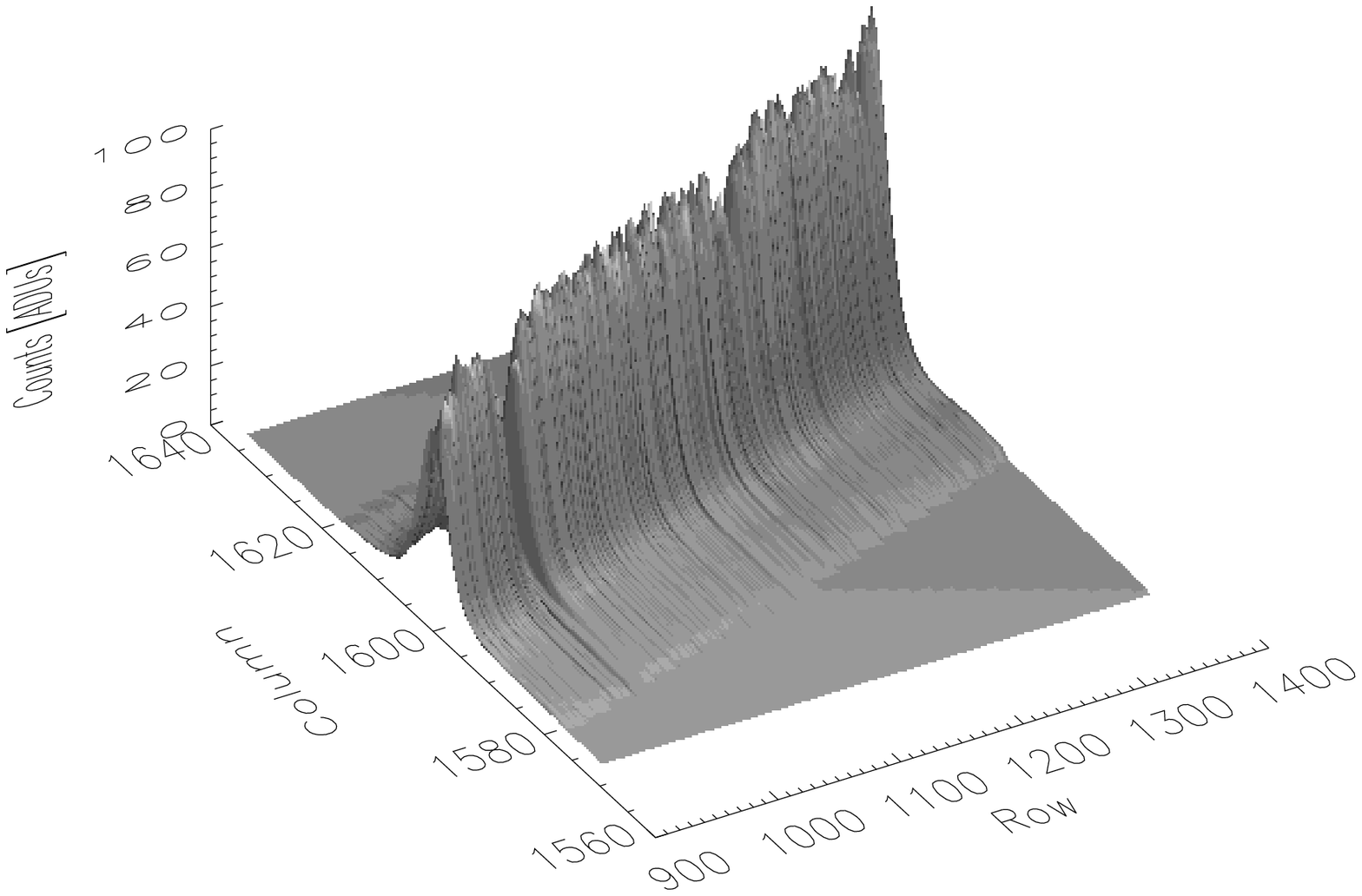}}
    \end{minipage}
    b)\hspace*{-4mm}
    \begin{minipage}[t]{0.33\textwidth}
      \centerline{\includegraphics[width=\textwidth]{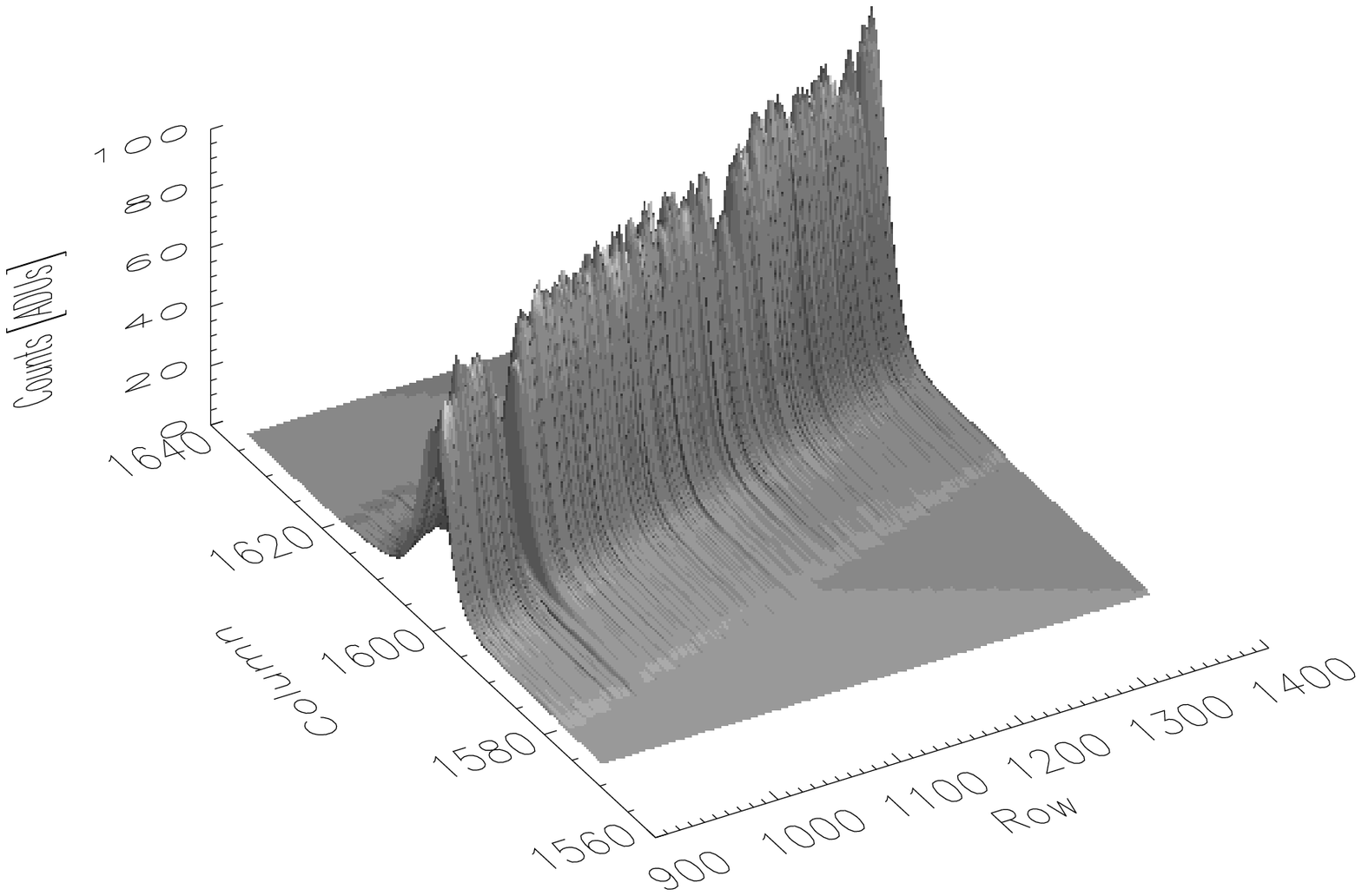}}
    \end{minipage}
    c)\hspace*{-4mm}
    \begin{minipage}[t]{0.33\textwidth}
      \centerline{\includegraphics[width=\textwidth]{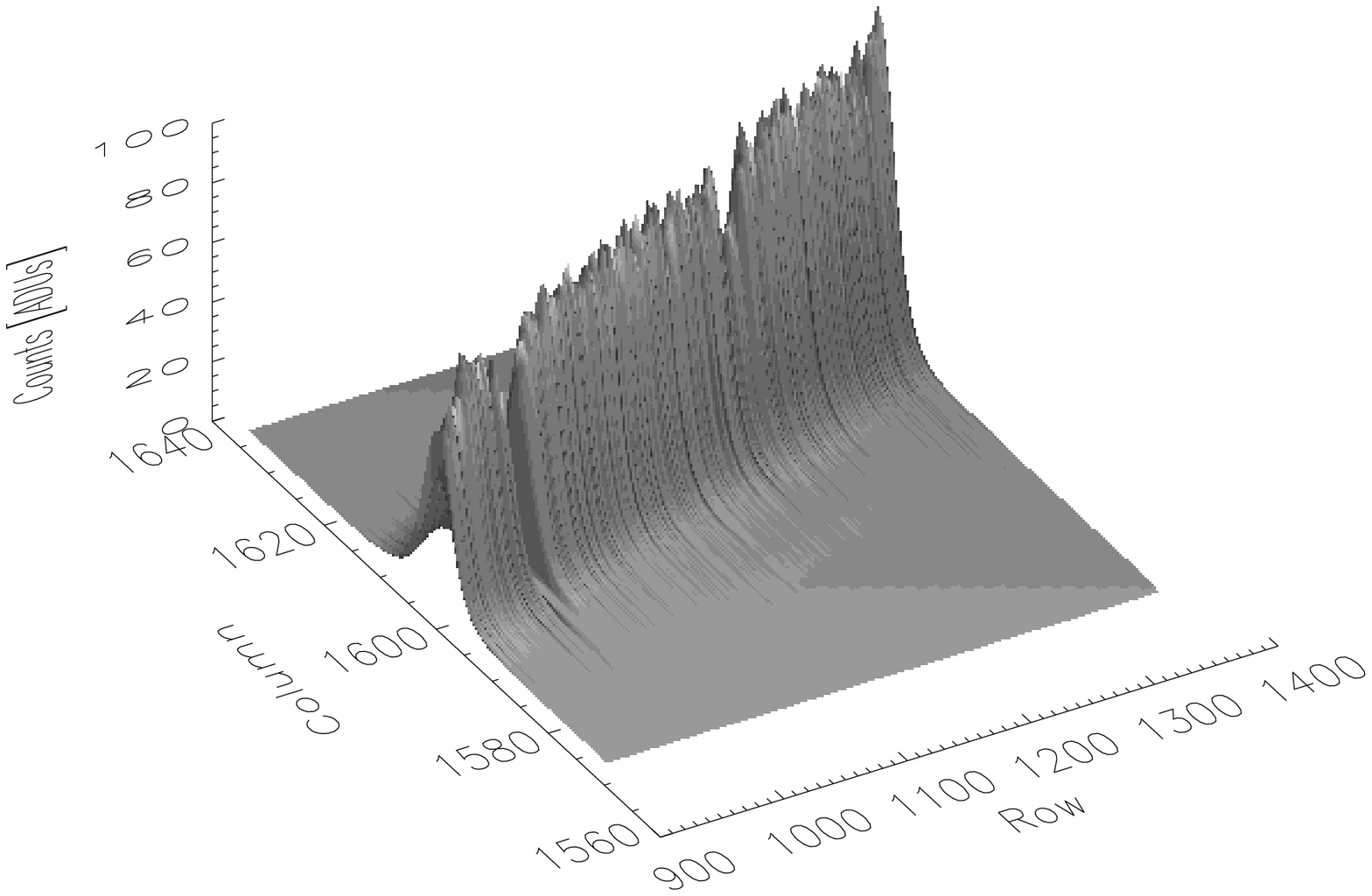}}
    \end{minipage}
  \end{minipage}
 \end{center}
 \caption{Comparison of the reconstructed object spectra from Fig.~\ref{fig:ShadeSurf_CCD_profiles}. a) Original
version by P\&V without proper cosmic-ray removal at row 1206. While the spike at the wing of the spatial profile is 
properly removed, the cosmic ray close to the profile center causes a spike in an absorption line in the extracted object 
spectrum, which partially fills in the absorption feature.\ b) Our re-implementation, all cosmic-ray hits properly removed.\ c) Our new algorithm with all spikes removed 
and the wings of the spatial profile going down to zero as they should.}
 \label{fig:ShadeSurf_reconstructed_object_spectra}
\end{figure*}
In Fig.~\ref{fig:ShadeSurf_reconstructed_object_spectra} we compare the reconstructed object spectra from the original 
version by P\&V (a), our re-implementation (b), and our new algorithm (c). While the cosmic-ray hit near the center of the 
spatial profile causes a spike at row 1206 in the extracted object spectrum shown in a), the cosmic rays are properly removed by our algorithms
shown in (b) and (c). The main difference between b) and c) is that in c) the profile goes down to zero at the wings (as one
would expect for a point source), while in b) too much weight is given to the background.

\begin{figure*}
 \begin{center}
  \begin{minipage}[t]{0.999\textwidth}
    \hspace*{-1mm}
    a)\hspace*{-4mm}
    \begin{minipage}[t]{0.33\textwidth}
      \centerline{\includegraphics[width=\textwidth]{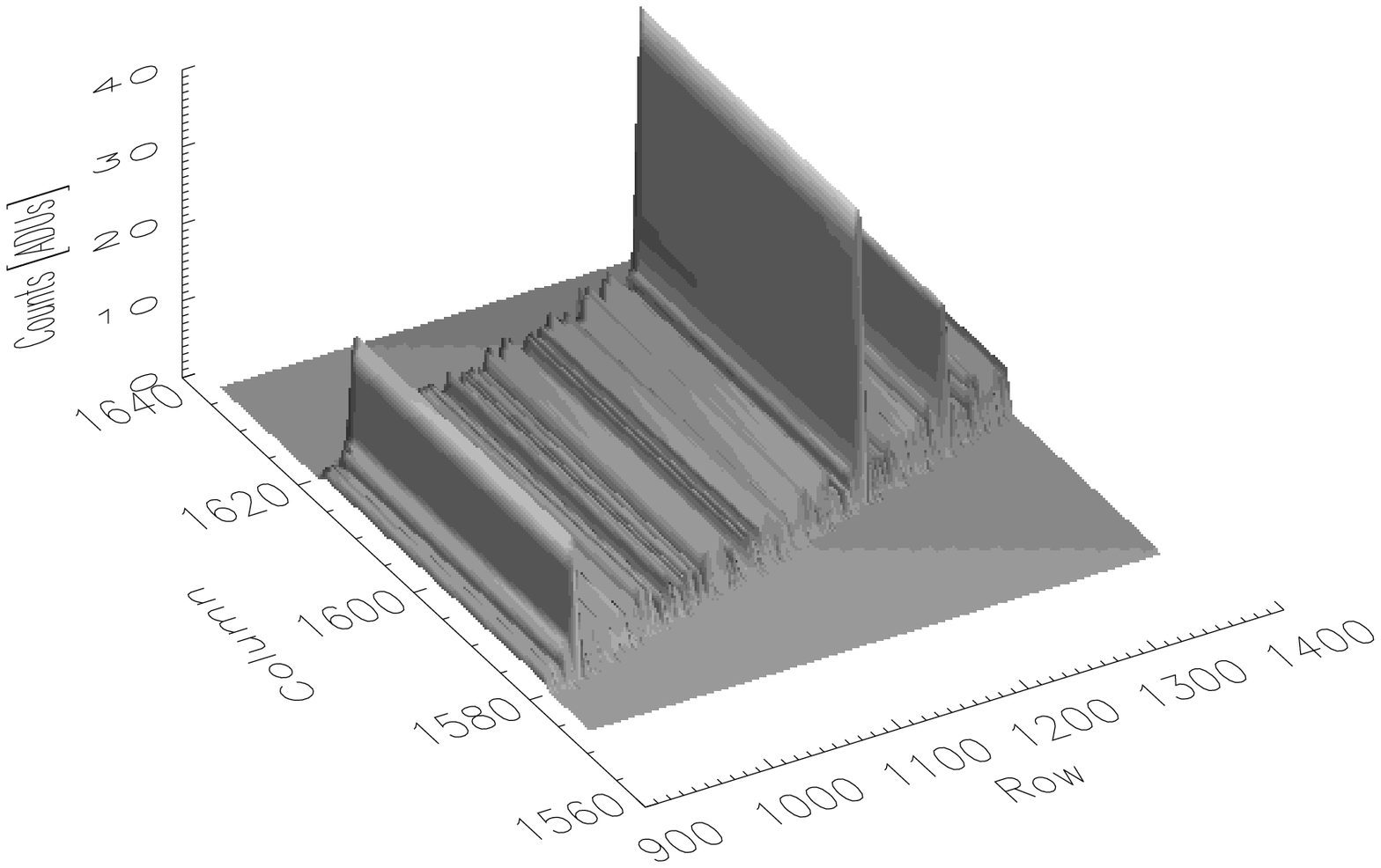}}
    \end{minipage}
    b)\hspace*{-4mm}
    \begin{minipage}[t]{0.33\textwidth}
      \centerline{\includegraphics[width=\textwidth]{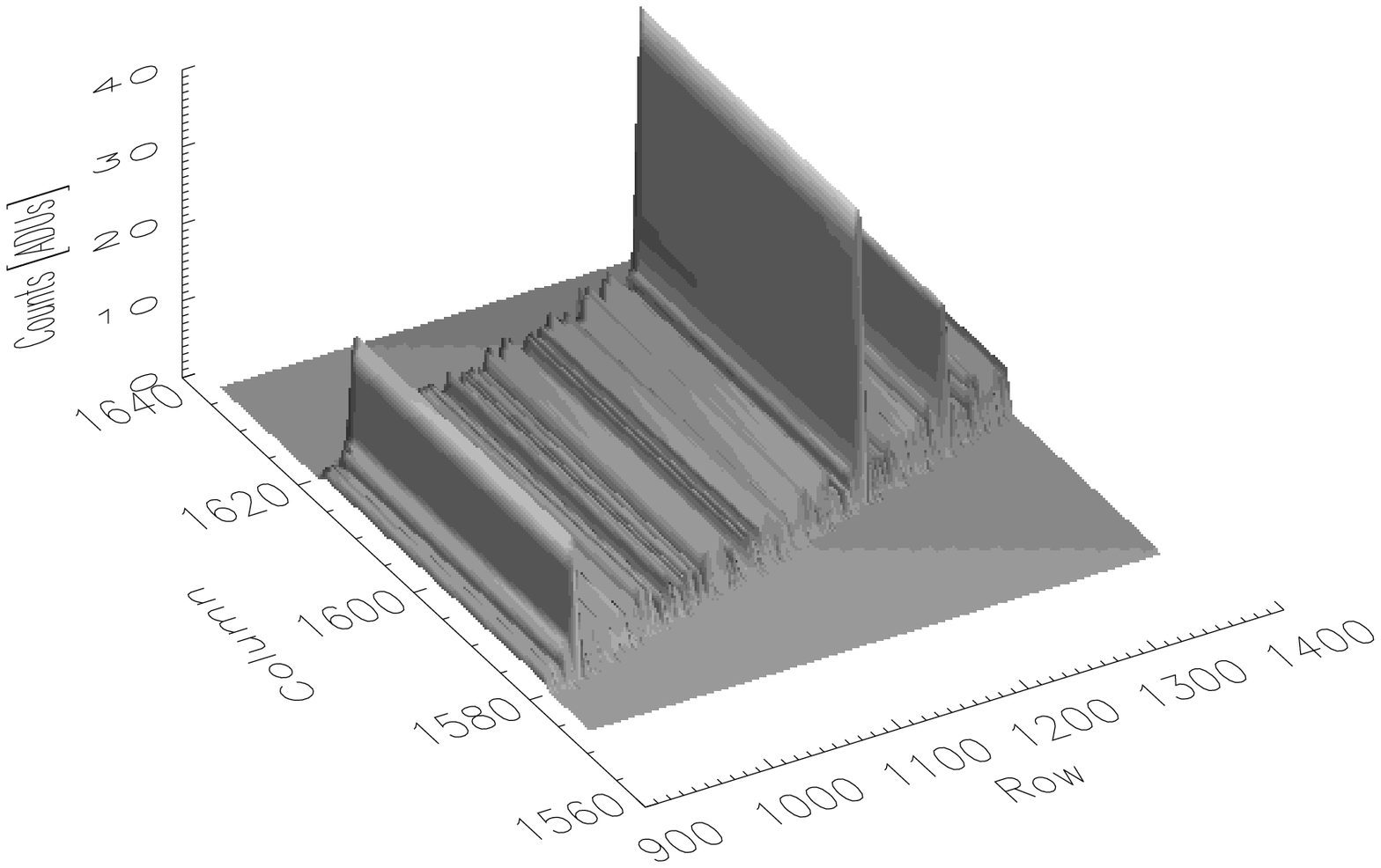}}
    \end{minipage}
    c)\hspace*{-4mm}
    \begin{minipage}[t]{0.33\textwidth}
      \centerline{\includegraphics[width=\textwidth]{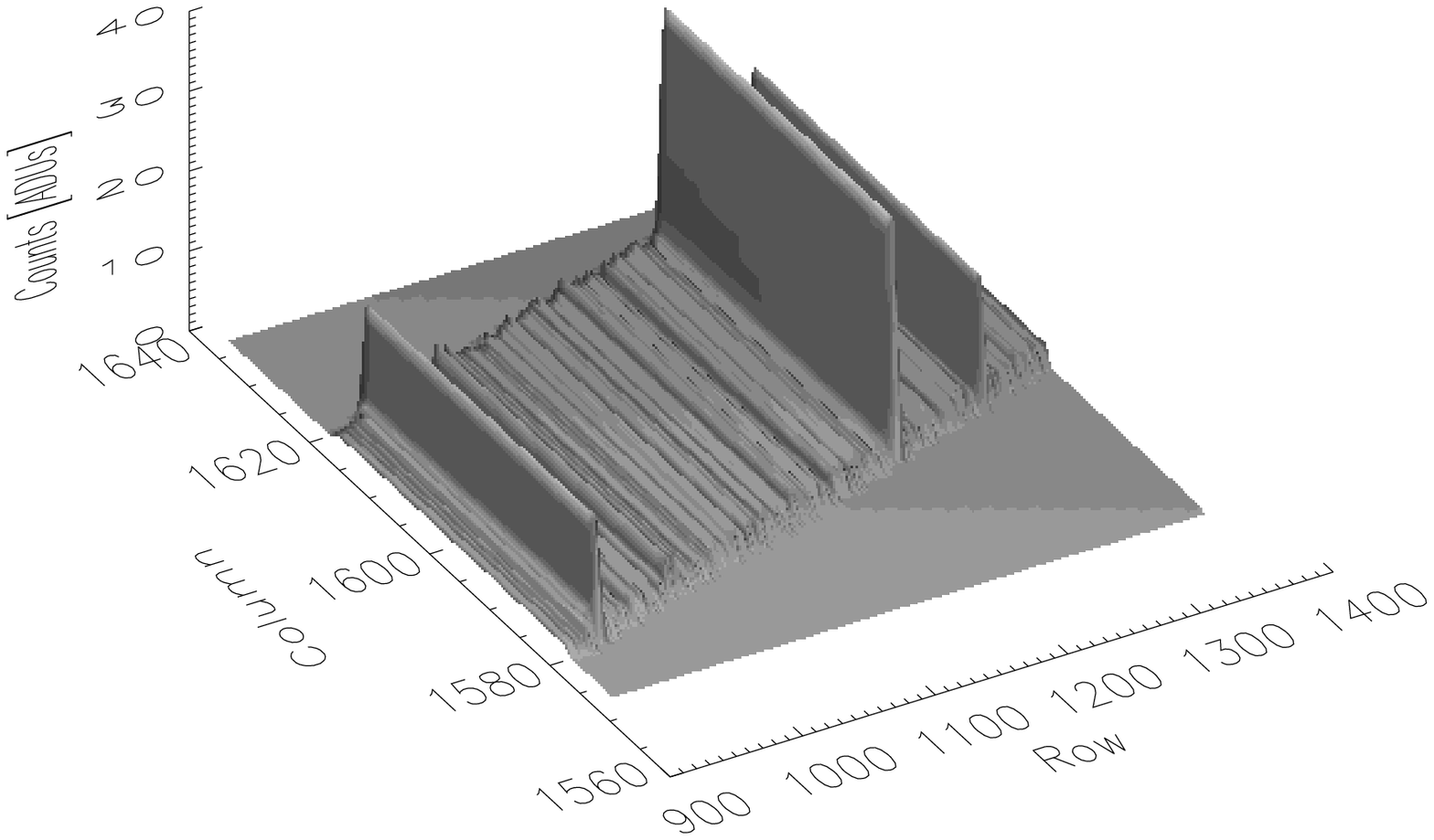}}
    \end{minipage}
  \end{minipage}
 \end{center}
 \caption{Comparison of the reconstructed sky spectra from Fig.~\ref{fig:ShadeSurf_CCD_profiles}. Visible are three sky 
 lines at rows 940 (5895.932 \AA), 1220 (5889.959 \AA), and 1300 (5888.192 \AA). The sky images are scaled the same and the 
 surrounding pixels are set to zero for a better visiblity of the sky.\ a) Original version by P\&V -- as expected nearly 
 identical to\ b) (our re-implementation). The sky line at row 1220 is slightly overestimated.\ c) Our new algorithm with the 
 smoothest sky and all sky lines properly reproduced (as shown below). Even an additional sky line at row 982 (5894.472 \AA) 
 clearly shows up which completely vanishes in the noise of the other reconstructed sky spectra.}
 \label{fig:ShadeSurf_reconstructed_sky_spectra}
\end{figure*}
The comparison of the reconstructed sky spectra in the cases of (a), (b), and (c) is shown in Fig.~\ref{fig:ShadeSurf_reconstructed_sky_spectra}. We have 
three sky lines visible at rows 940 (5895.932 \AA), 1220 (5889.959 \AA), and 1300 (5888.192 \AA). In (a) the sky is relatively noisy and the strongest line 
is 13\% higher than in our (c) method.
As expected, our re-implementation of the original algorithm leads to nearly identical results (b). Our new optimal 
extraction and sky-subtraction algorithm (c) leads to the smoothest sky results. All sky lines are properly 
reproduced, as shown below. Using our new algorithm we additionally find a new sky line at row 982. This line can only be
inferred in the other sky spectra. A comparison to the catalogue of optical sky emission from
UVES (Hanuschik 2003) identifies it as the 5894.472 \AA~line.

\begin{figure*}
 \begin{center}
  \begin{minipage}[t]{0.999\textwidth}
    \hspace*{-1mm}
    a)\hspace*{-4mm}
    \begin{minipage}[t]{0.33\textwidth}
      \centerline{\includegraphics[width=\textwidth]{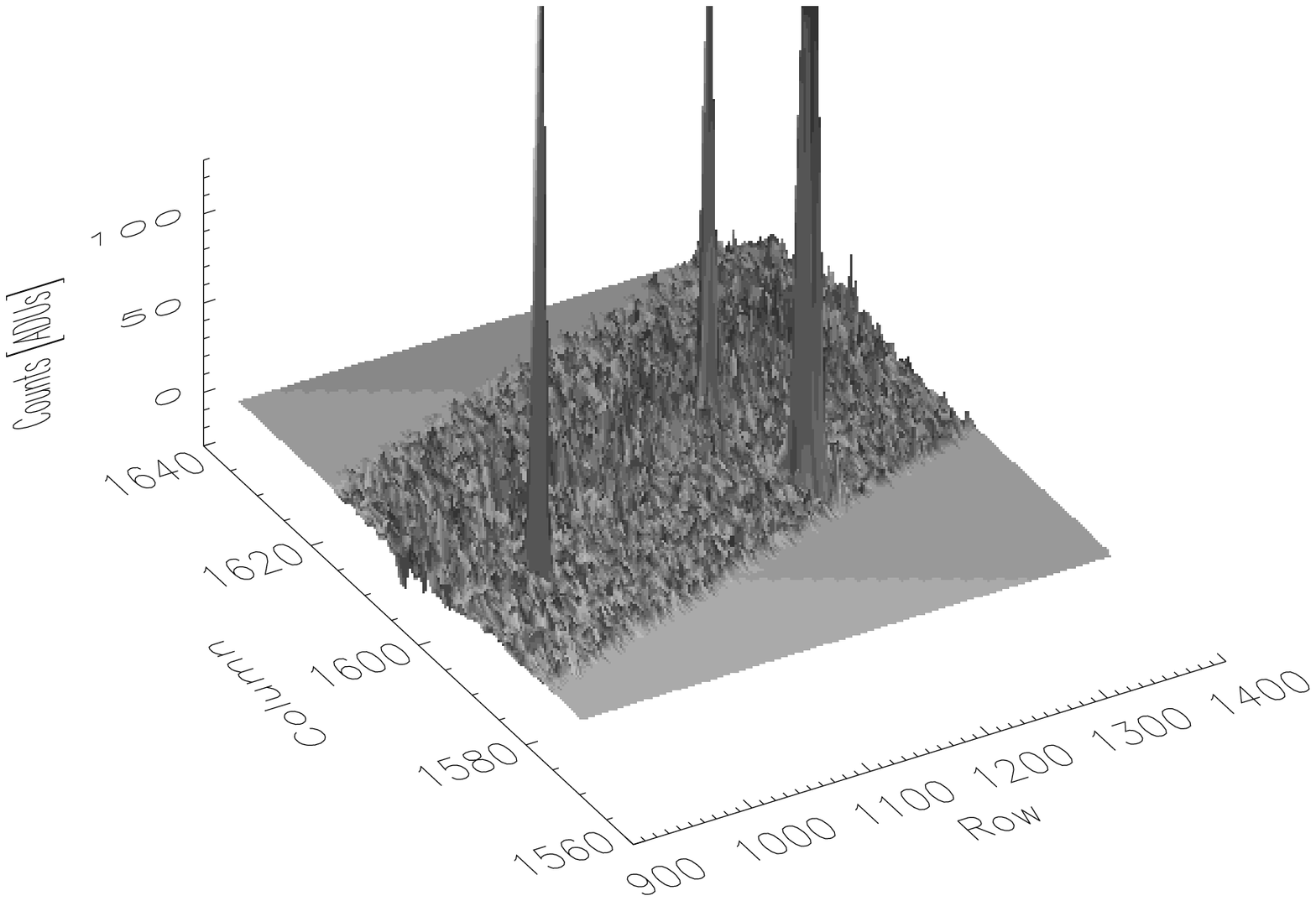}}
    \end{minipage}
    b)\hspace*{-4mm}
    \begin{minipage}[t]{0.33\textwidth}
      \centerline{\includegraphics[width=\textwidth]{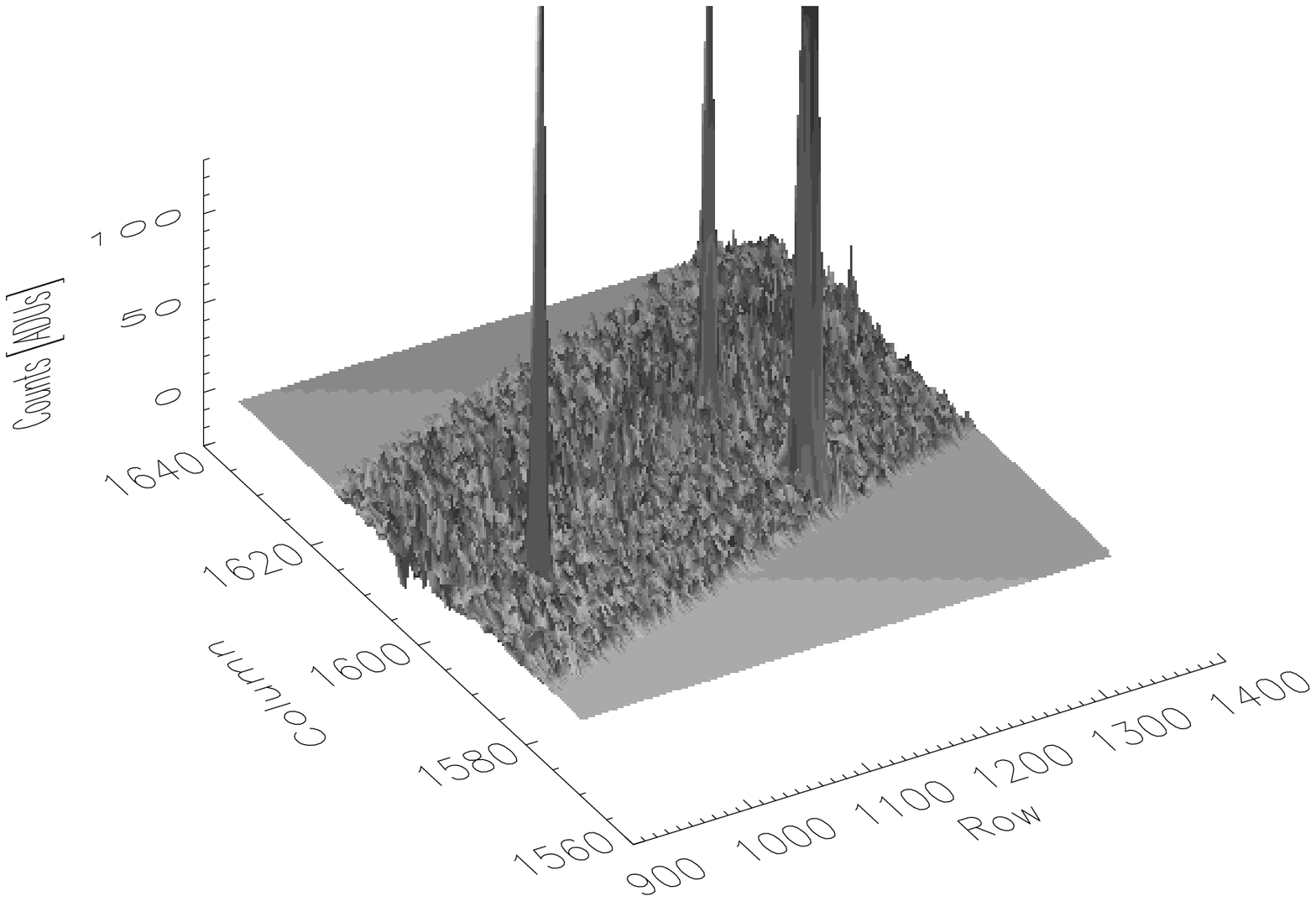}}
    \end{minipage}
    c)\hspace*{-4mm}
    \begin{minipage}[t]{0.33\textwidth}
      \centerline{\includegraphics[width=\textwidth]{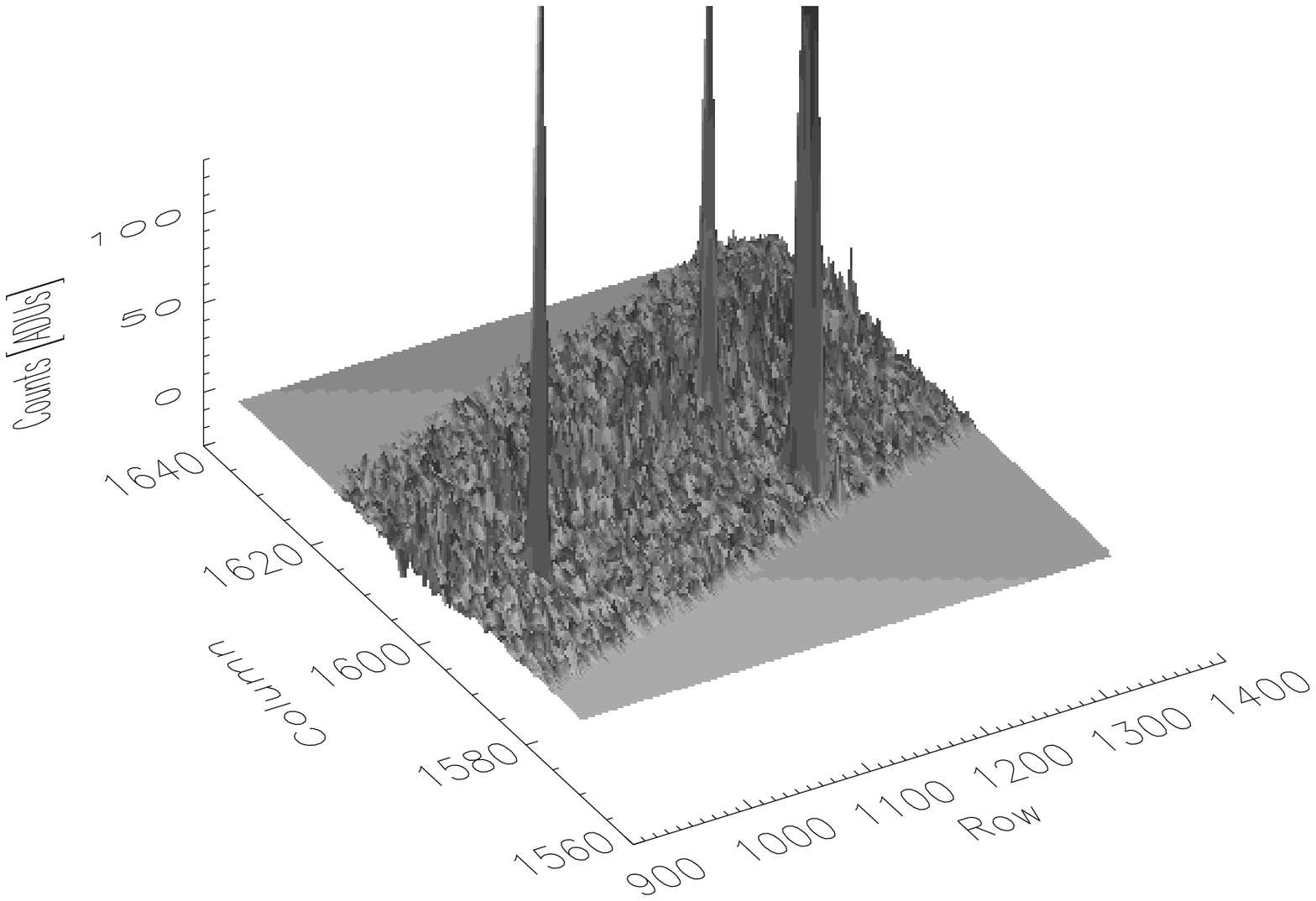}}
    \end{minipage}
  \end{minipage}
 \end{center}
 \caption{Comparison of the differences between the CCD image from Fig.~\ref{fig:ShadeSurf_CCD_profiles} and the 
 reconstructed object and sky spectra from Fig.s~\ref{fig:ShadeSurf_reconstructed_object_spectra} 
 and~\ref{fig:ShadeSurf_reconstructed_sky_spectra}. No obvious remaining structures which would indicate systematic discrepancies between the 
 CCD image and the reconstruction can be seen.}
 \label{fig:ShadeSurf_diff_orig_reconstructed}
\end{figure*}
The differences between the CCD image shown in Fig.~\ref{fig:ShadeSurf_CCD_profiles} and the reconstructed object and sky
spectra from Fig.s~\ref{fig:ShadeSurf_reconstructed_object_spectra} 
and~\ref{fig:ShadeSurf_reconstructed_sky_spectra} are shown in Fig.~\ref{fig:ShadeSurf_diff_orig_reconstructed}. No remaining 
structures which would indicate systematic discrepancies between the CCD image and the reconstruction are obvious in any of the images.

\begin{figure}[t]
  \plotone{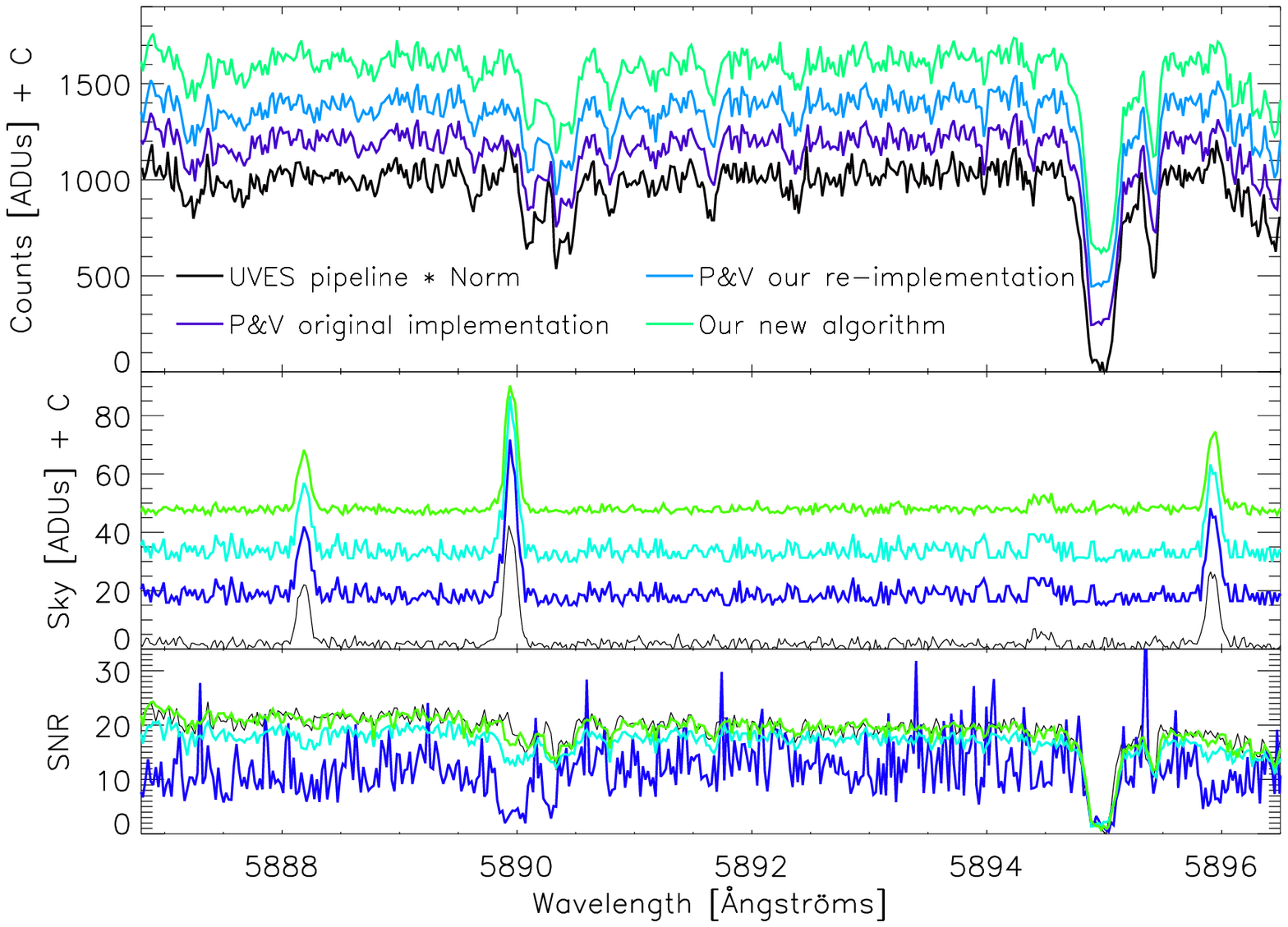}
  \caption{Plot comparing the resulting object and sky spectra of the spectral region shown in
  Fig.~\ref{fig:ShadeSurf_CCD_profiles}, extracted with the original sky-subtraction algorithm from
  P\&V, our new programs, and the UVES pipeline. The upper part of the image shows the
  extracted object spectra, the middle part shows the extracted sky, and the lower part shows the estimated SNR. In the 
  upper two parts the spectra were shifted upwards for better visibility. Note the spike at $5890.29~\mathrm{\AA}$ in the 
  UVES spectrum and the original implementation by P\&V, which is caused by a cosmic-ray hit.}
\label{fig:ComparisonSkySTELLA-UVES}
\end{figure}

\begin{figure}[t]
  \centerline{\includegraphics[width=0.1\textwidth]{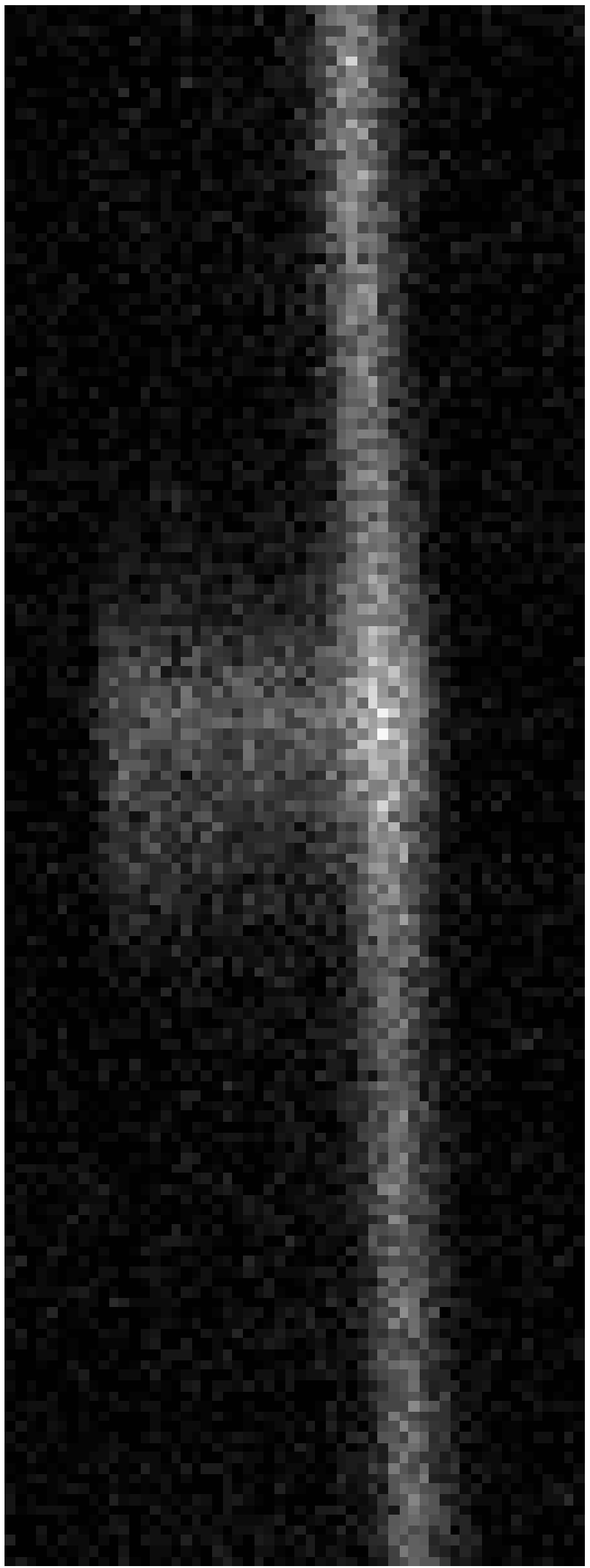}\hfill
  \includegraphics[width=0.1\textwidth]{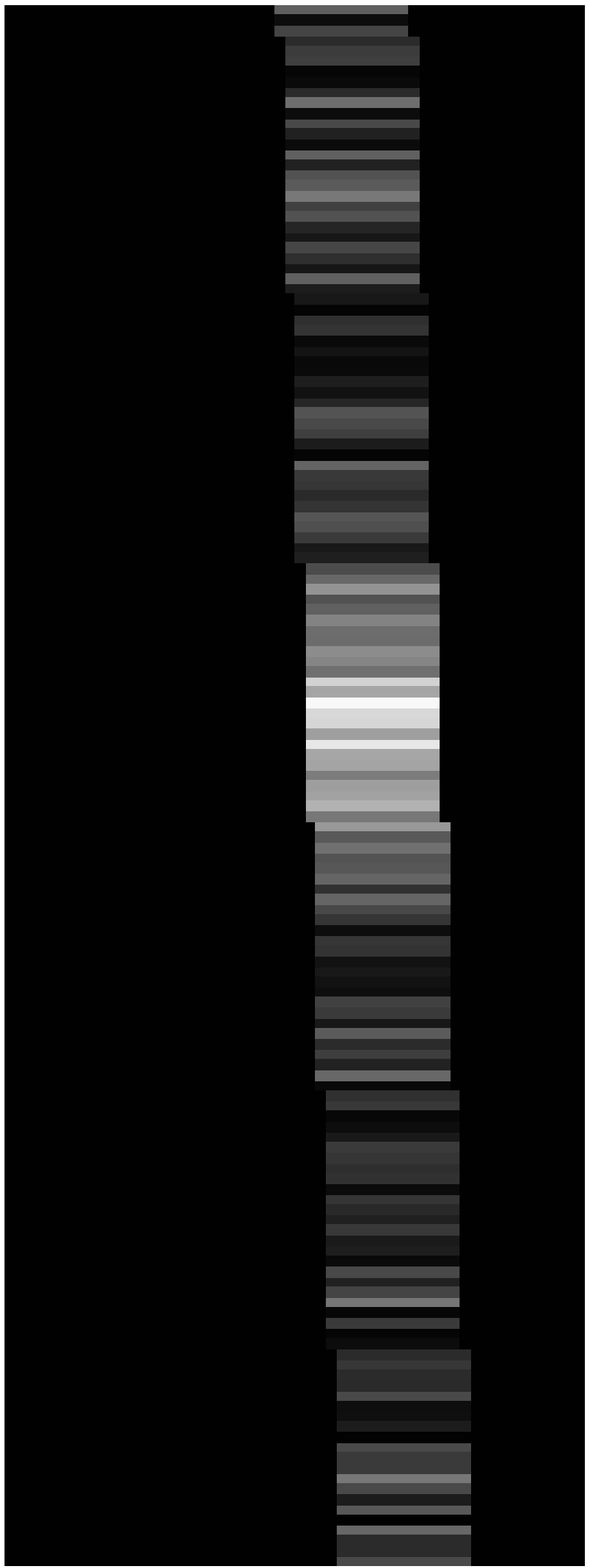}\includegraphics[width=0.1\textwidth]{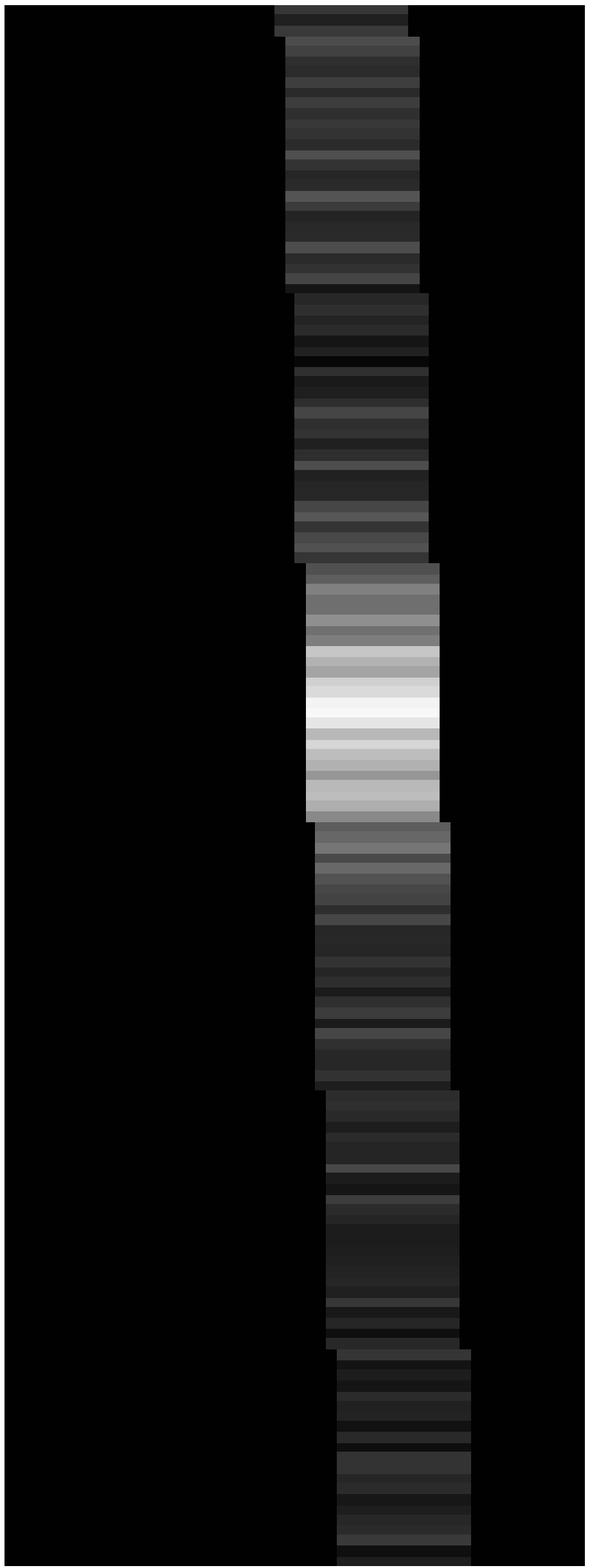}\hfill
  \includegraphics[width=0.1\textwidth]{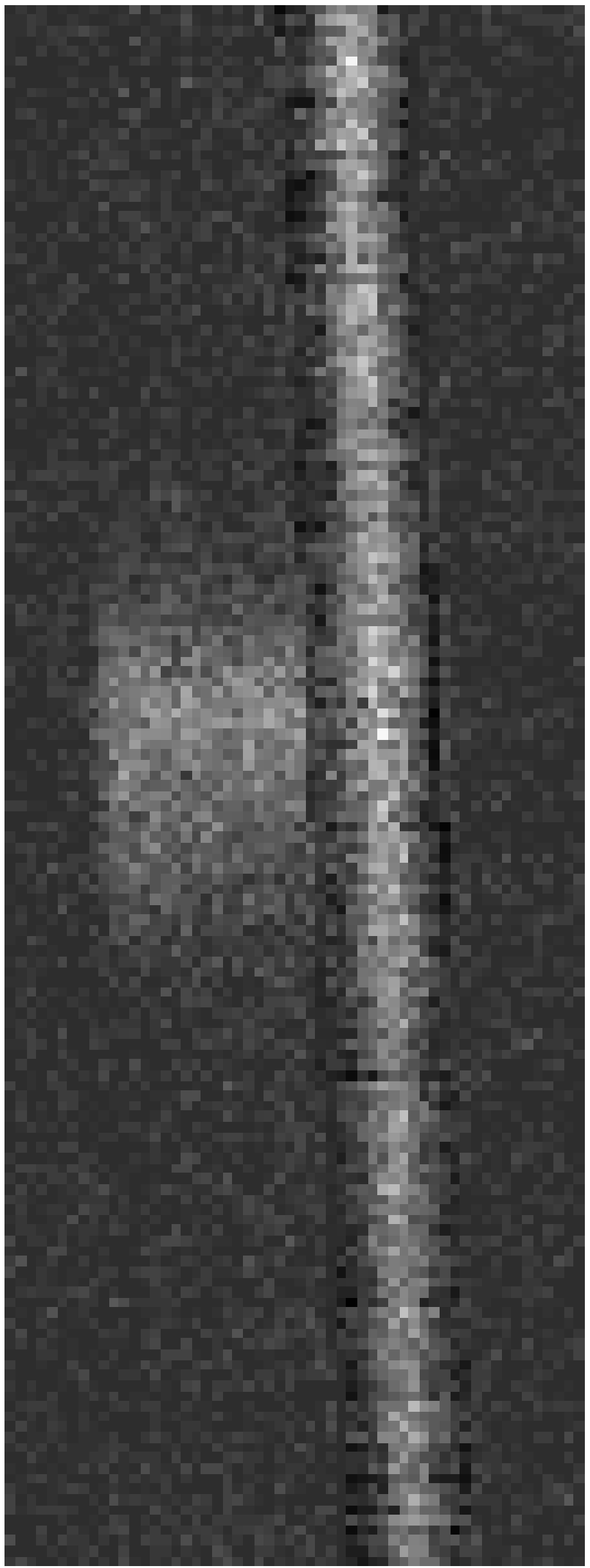}\includegraphics[width=0.1\textwidth]{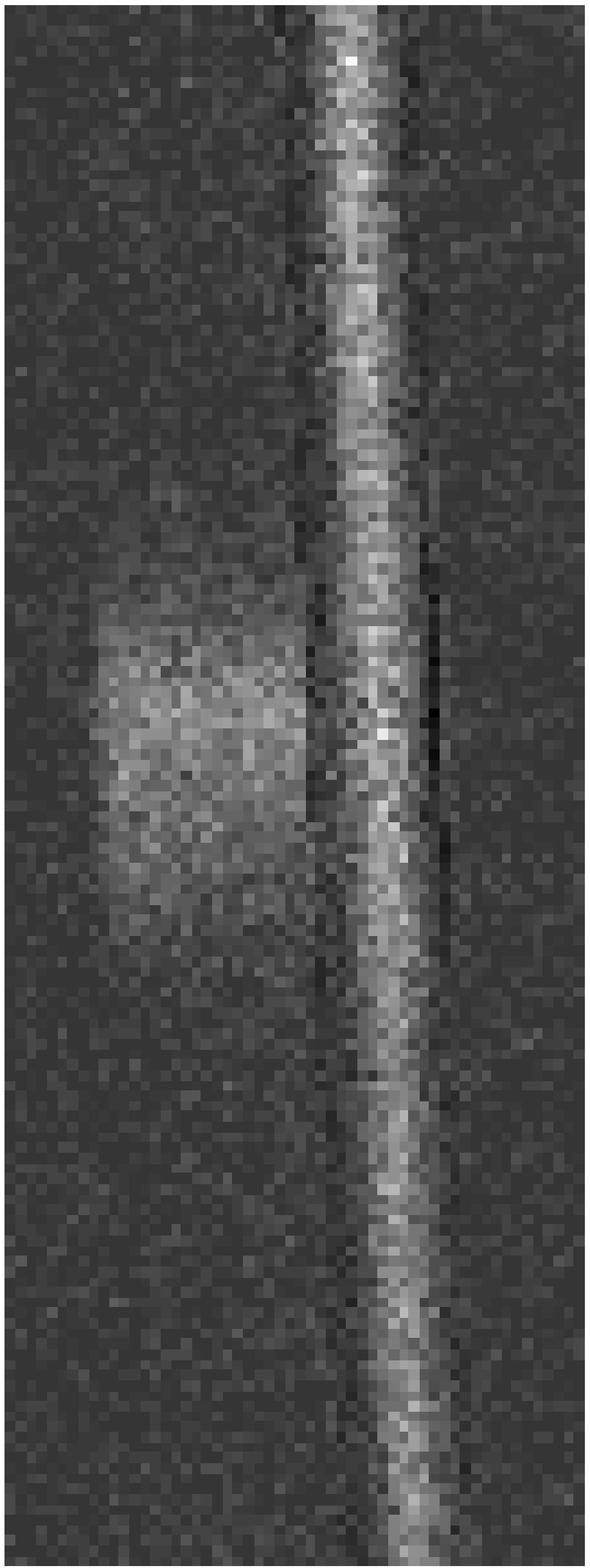}\hfill
  \includegraphics[width=0.1\textwidth]{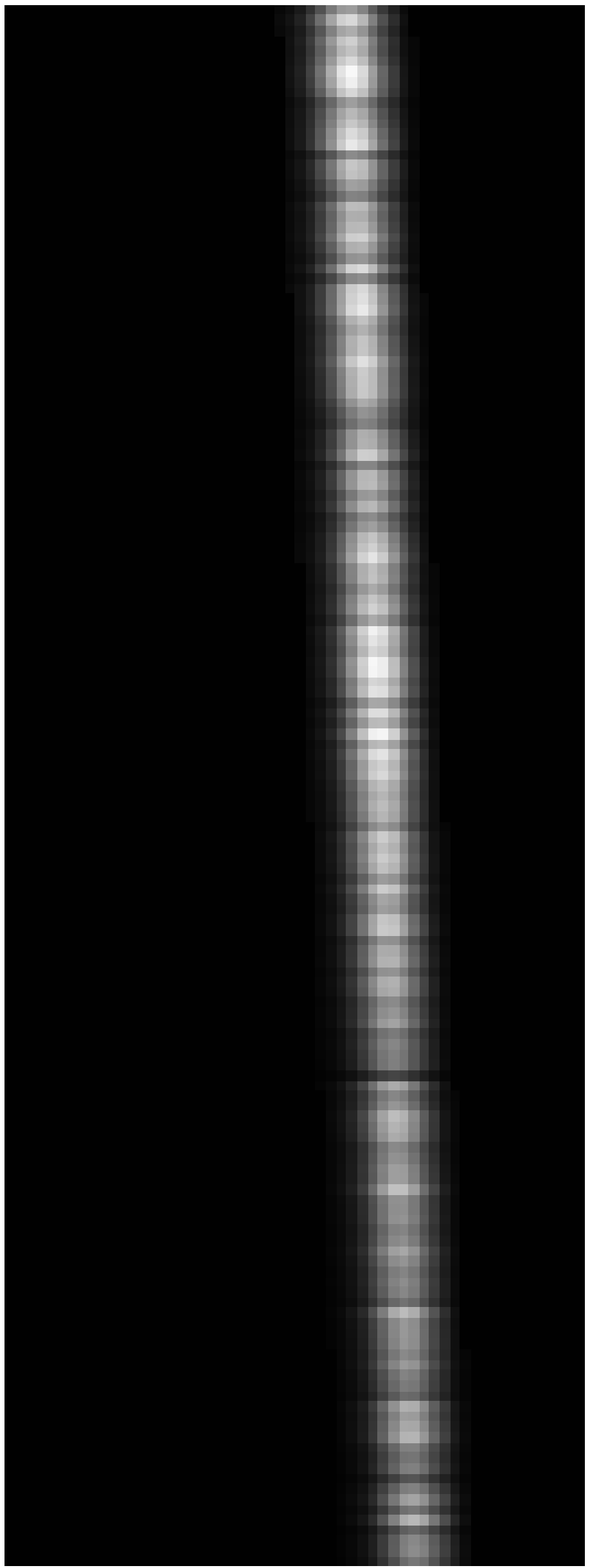}\includegraphics[width=0.1\textwidth]{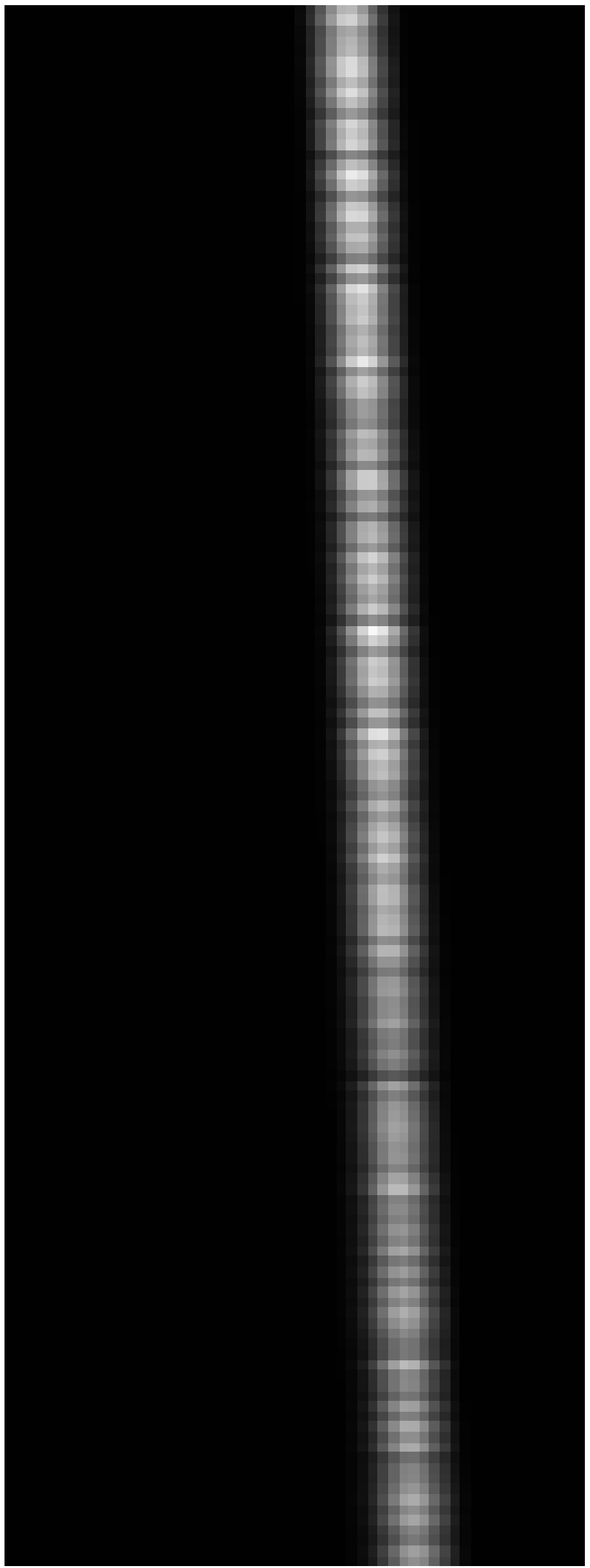}\hfill
  \includegraphics[width=0.1\textwidth]{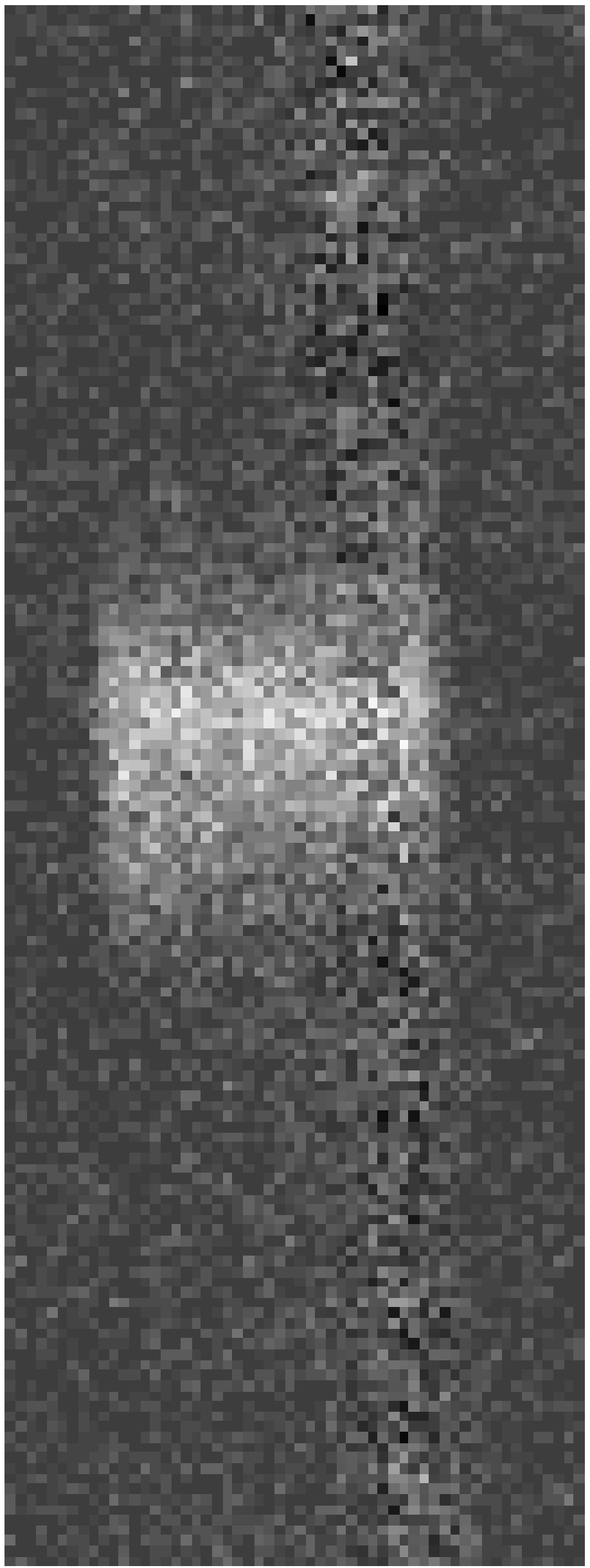}\includegraphics[width=0.1\textwidth]{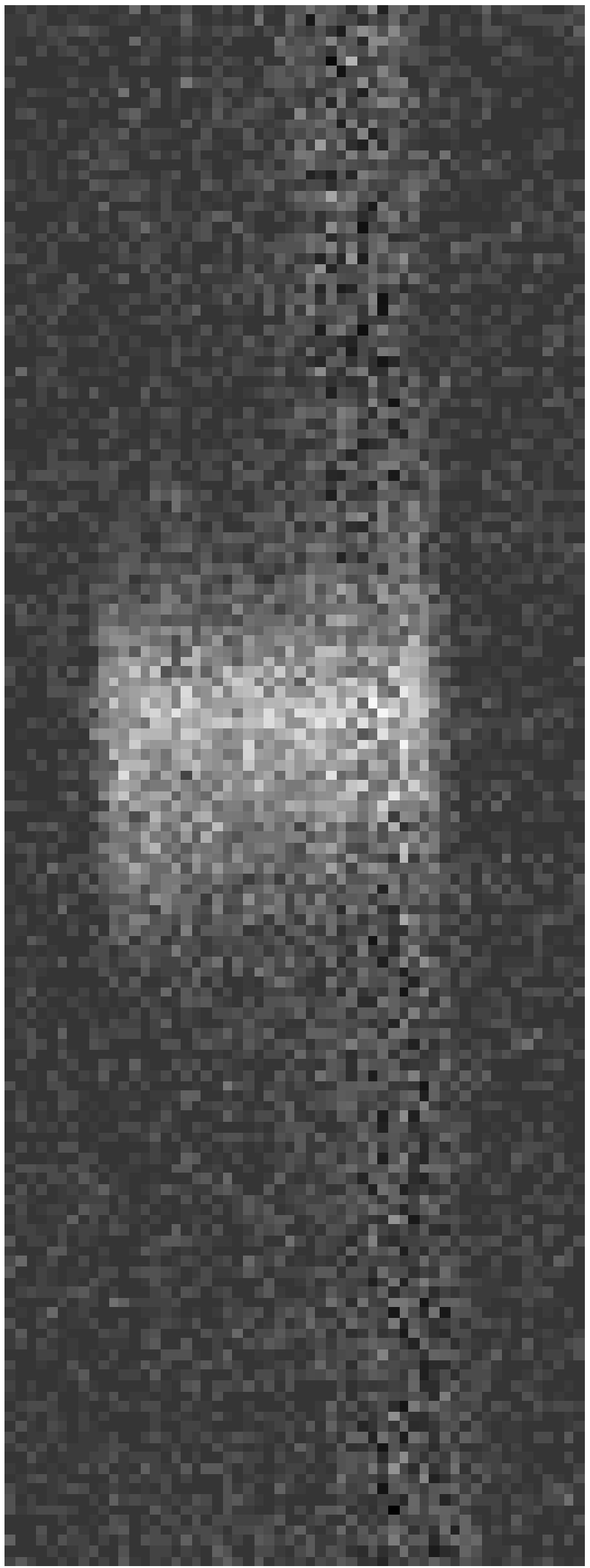}}
  \caption{Plot comparing the resulting sky (actually accretion disk) and object spectra for a short part of an UVES spectrum of LMC-X1 in the blue arm, 
  extracted with our re-implementation of the original sky-subtraction algorithm from P\&V and our new programs. The most left image is the bias and scattered-light
  subtracted and flat-fielded CCD image, followed by the reconstructed sky (original algorithm and our new programs), CCD minus reconstructed sky, reconstructed object 
  image, and CCD minus reconstructed object. As the object is close to the slit edge, the original algorithm by P\&V is struggling to remove the sky properly, while our
  new programs deliver a much better result. Note that the apparent over subtraction of the sky in the middle panel is not real and only due to the fact that the right 
  aperture limit is outside the slit's image where there is no signal.}
\label{fig:ComparisonSkyEdgeDS9}
\end{figure}

\begin{figure}[t]
  \plotone{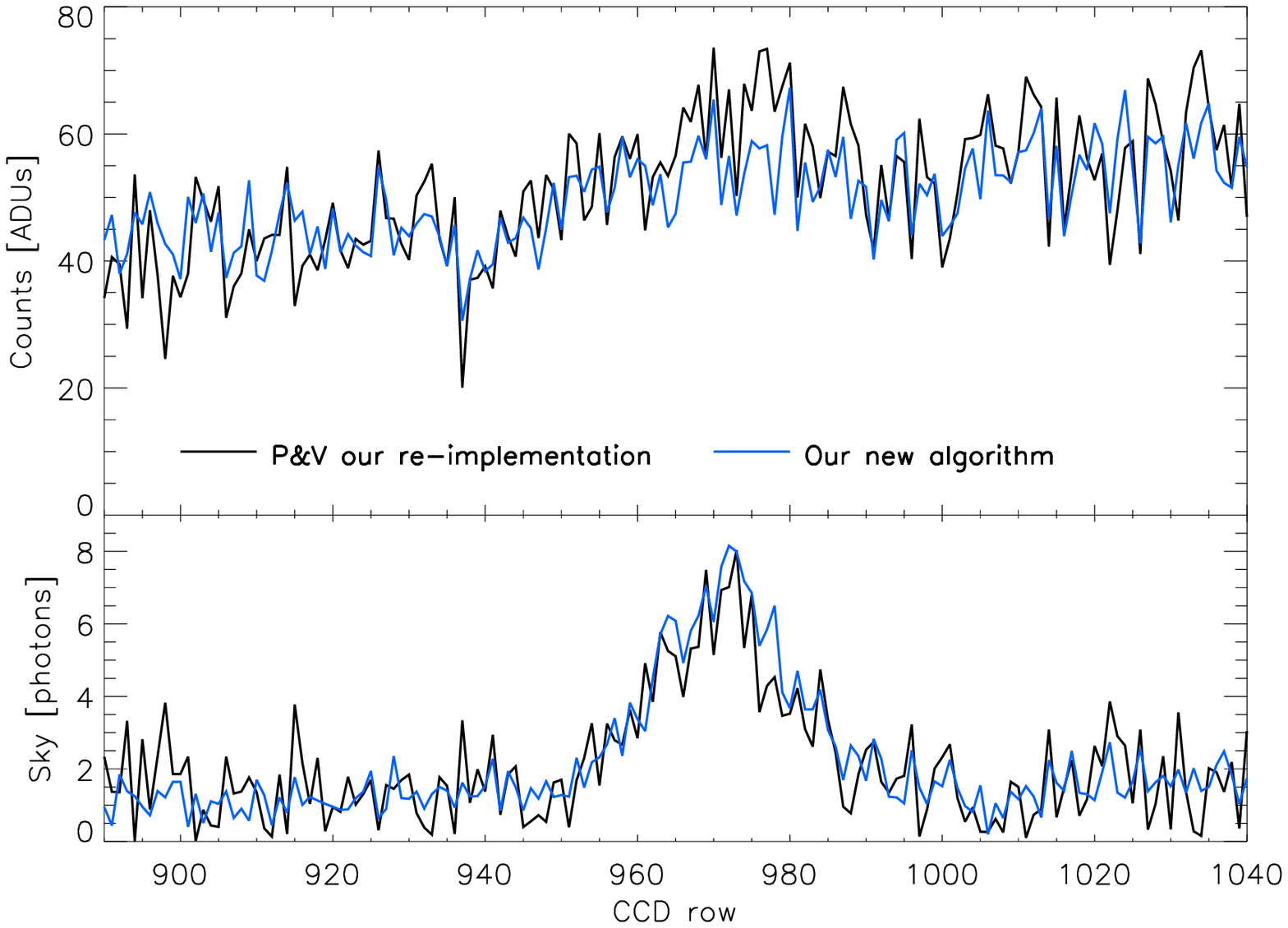}
  \caption{Plot of the object and sky spectra shown in Fig.\,\ref{fig:ComparisonSkyEdgeDS9}. The SNR measured with IRAF splot is 6.3 for the original algorithm by P\&V
  and 9.1 for our new algorithms -- an improvement of nearly 50\%.}
\label{fig:ComparisonSkyEdge}
\end{figure}

The comparison of the resulting sky and object spectra from the original implementation by P\&V to our re-implementation, 
our new algorithms, as well as to the UVES pipeline, is presented in
Fig.~\ref{fig:ComparisonSkySTELLA-UVES}.
Shown are the optimally extracted object spectra, the sky spectra, and the 
estimated SNRs. 
The UVES pipeline uses a similar approach as we present here, but always assumes a Gaussian for the spatial
profile, leading to systematic errors. The sky is calculated by fitting the Gaussian profile plus constant (sky) to the 
CCD row. Again, assuming a Gaussian for the spatial profile is leading to systematic errors in the sky as well. As it can
be seen in the figure, our new algorithms are leading to the smoothest sky and the least sky- and cosmic-ray residuals in the 
object spectrum. As expected, the sky values calculated with the REDUCE pipeline and our re-implementation are nearly identical.
Note however that the REDUCE pipeline, as well as the UVES pipeline, produce a spike at $5890.29~\mathrm{\AA}$ caused by a 
residual cosmic-ray hit. The SNR calculated by the REDUCE pipeline is unrealistic (we still cannot reproduce 
Eq.~\ref{eq:reduce_sigma}). In contrast, our error estimates (Eq.~\ref{eq:sigma_orig} for our re-implementation and Eq.s~\ref{eq:sigma_s_new}
and \ref{eq:sigma_b_new} for our new algorithms), using the propagated uncertainties for every pixel, are similar to the 
error estimates stated by the UVES pipeline. However, as we could not find a documentation on the algorithm used to calculate
the UVES uncertainties, we cannot comment on the stated UVES SNR too much. As the object extraction 
and sky subtraction are done in a way very similar to our new algorithms, we can assume that the errors are also calculated
in a similar way. Given that the spatial profile is close to a Gaussian, this explains why the SNR stated by the UVES pipeline
is very close to the SNR for our new algorithms. Note however that the UVES object
spectrum was rebinned using an oversampling factor of $\sim 1.5$. While oversampling preserves the resolution, it smooths the 
spectrum, making it appear as if it had a higher SNR. Visibly, our new extraction method leads to an object spectrum even 
smoother than the rebinned spectrum produced by the UVES pipeline. Compared to the SNR for the original algorithm by P\&V, 
calculated from the propagated pixel uncertainties, our new algorithms lead to a SNR $\approx 15-20$ \% higher. The 
simple-sum extraction (which is not shown in the plot because of the sky lines and cosmic-ray hits) leads to a calculated 
SNR of only $\approx 3.5$, even with an aperture width only slightly larger than the object spectrum!\\
In Fig.s\,\ref{fig:ComparisonSkyEdgeDS9} and \ref{fig:ComparisonSkyEdge} a comparison of the performance of the original sky-subtraction algorithm by P\&V to our new 
programs is shown for the case that the object is close to the slit edge. The images show a short part of a spectrum of LMC-X1, taken with UVES in the blue arm. While 
the original sky-subtraction algorithm by P\&V is struggling to remove the background properly, our new algorithm performs much better, resulting in a $\approx50\%$ gain
for the SNR of the object spectrum. Note that the spectrum actually shows an accretion disk surrounding the black hole and not a sky emission line. 

\begin{figure}
  \plotone{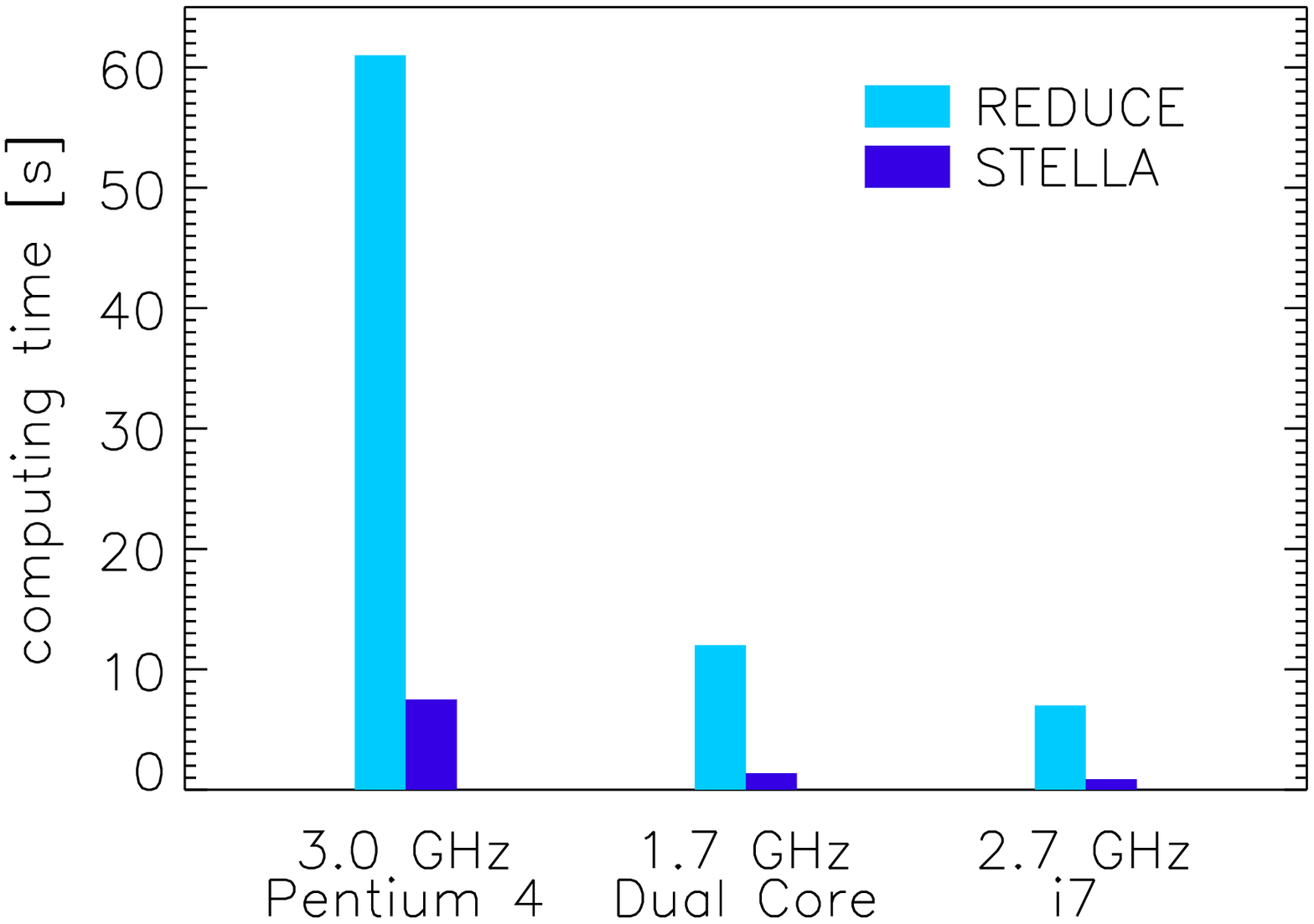}
  \caption{Plot comparing the computing time for the calculation of the aperture profile and optimal extraction
  of one order (15x4089 pixels) for the ESO/FEROS spectrograph. Our \texttt{C/C++} 
  re-implementation of P\&V's algorithms is about 10 times faster than the original \texttt{IDL} procedures, making it 
  perfectly suitable for large spectroscopic surveys.}
\label{fig:Performance}
\end{figure}

Shown in Fig.~\ref{fig:Performance} is the computing time of our new programs
and the original REDUCE procedures for one spectral order of the ESO/FEROS spectrograph
(18x4089 pixels). It shows that our \texttt{C/C++} re-implementation is about 10 times
faster than the \texttt{IDL} version. In times of major spectroscopic surveys with
multi-object spectrographs, this is a major advantage over the original programs.\\

Our new programs have been extensively tested with a wide range of spectrographs, including slits and fibre feeds, prisms and 
gratings, Echelle spectra, multi-object spectra, and IFUs. Comparisons of the STELLA
pipeline to the results of the pipelines provided by the individual observatories and
to standard IRAF have shown that our pipeline can lead to significant SNR improvements. 
The UVES pipeline has been
optimised for low SNRs and always assumes a Gaussian profile perpendicular to the
dispersion axis for the optimal-extraction algorithm. For low SNR the random error of the
Poisson noise (photon noise) is larger than the systematic error introduced by this
approximation. In the low-SNR regime, assuming a Gaussian spatial profile is therefore an
acceptable approximation. However, profile determination is the most critical step for variance
weighted (optimal) extraction. Assuming an incorrect spatial profile can lead to large
systematic errors for high-SNR data. These systematics drop the achievable SNR significantly
and can introduce ripples in the extracted spectra, both leading to larger uncertainties in the
stellar parameters derived from these spectra.\\
The optimal-extraction algorithms of standard IRAF are also known to have room for improvement.
The simple sum nearly always leads to a higher SNR than these `optimal extraction' algorithms,
even if \texttt{apsum} is fed with the most probable profile calculated with P\&V's
profile-fitting algorithm. Our new algorithms are now allowing for the utilisation of a state-of-the-art optimal-extraction 
algorithm within the IRAF environment, as well as most other existing data-reduction packages.\\

\section{Conclusions}
\label{sec:Conclusion}
The presented new implementation of the state-of-the-art optimal-extraction algorithm developed
by P\&V can now easily be integrated in IRAF and most other existing data-reduction packages and
programming languages.
It allows for scattered-light subtraction, profile calculation, error propagation, and optimal
extraction of Coud\'e and Echelle slit spectra as well as fibre feeds.
For slit spectra a new optimal sky ex-/subtraction algorithm at the Poisson limit was developed,
which works even for extended objects, short slits, or if the observed star is at the slit's
end.
Our progams offer the new optimal extraction and sky-subtraction algorithms as well as the
original ones by P\&V. The re-implementation in C++ is about 10 times faster than the original one, making it
perfectly suitable for large surveys.
Our new programs have been extensively bug fixed and successfully tested with spectra from VLT/UVES, ESO/CES,
ESO/FEROS, NTT/EMMI, NOT/ALFOSC, STELLA/SES, SSO/WiFeS, and, finally P60/SEDM-IFU. 
As already shown by P\&V, their optimal-extraction algorithm can lead to a significant improvement in the 
SNR compared to standard IRAF and other DRPs. In this paper we have shown that
using our improved algorithms for cosmic-ray/CCD-defect removal and background subtraction we can 
again achieve a SNR gain of 15-50\% compared to the original algorithms by P\&V.
This is leading to much smaller errors in parameter estimates calculated from our optimally 
extracted spectra compared to other DRPs.

\section{Acknowledgements}
\label{sec:Acknowledgements}

We would like to thank the anonymous referee for the helpful comments, which significantly increased the scientific content
of this paper, and Heidi Viets and Andrew Fathy for proof reading the manuscript. Funding for the project has been provided 
by the Taiwanese National Science Council (grant numbers NSC 101-2112-M-008-017-MY3 and NSC 101-2119-M-008-007-MY3).

\newpage




\label{lastpage}

\end{document}